\documentclass[11pt,a4paper]{iopart}
\usepackage{graphicx,color}
\usepackage{dcolumn}
\usepackage{epsfig}
\usepackage{setstack}

\usepackage{iopams}
\usepackage{euscript}
\usepackage{amssymb}
\usepackage{amsthm}
\usepackage{newlfont}
\usepackage{mathrsfs}
\usepackage{subfigure}
\usepackage{todonotes,cite}
\usepackage{subfigure}
\usepackage[colorlinks,linkcolor=blue,urlcolor=blue,citecolor=blue]{hyperref}
\usepackage{ulem}

\newcommand{\opunit}{\textrm{1}\kern-0.22em\textrm{l}}

\usepackage{makeidx}

\def\bea{\begin{eqnarray}}
\def\eea{\end{eqnarray}}
\def\ba{\begin{array}}
\def\ea{\end{array}}

\def\pon{P_{\text{on}}}
\def\pof{P_{\text{off}}}
\def\hon{\hat{P}_{\text{on}}}
\def\hof{\hat{P}_{\text{off}}}

\newcommand{\aref}[1]{\ref{#1}}%

\def\bea{\begin{eqnarray}}
\def\eea{\end{eqnarray}}
\def\ba{\begin{array}}
\def\ea{\end{array}}

\begin{document}
\title{Brownian motion under intermittent harmonic potentials}
\author{Ion Santra $^{1^*}$, Santanu Das $^{2^*}$, Sujit Kumar Nath $^{3^*}$}
\address{$^1$ Raman Research Institute, Bengaluru 560080, India\\ $^{2}$ International Centre for Theoretical Sciences, Tata Institute of Fundamental Research, Bengaluru 560089, India \\$^3$ School of Computing and Faculty of Biological Sciences, University of Leeds, Leeds LS29JT, UK\\
$^*$ The authors have equal contributions.}

\begin{abstract}
We study the effects of an intermittent harmonic potential of strength $\mu = \mu_0 \nu$---that switches on and off stochastically at a constant rate $\gamma$, on an overdamped Brownian particle with damping coefficient $\nu$. This can be thought of as a realistic model for realisation of stochastic resetting. We show that this dynamics admits a stationary solution in all parameter regimes and compute the full time dependent variance for the position distribution and find the characteristic relaxation time. We find the exact non-equilibrium stationary state distributions in the limits---(i) $\gamma\ll\mu_0 $ which shows a non-trivial distribution, in addition as $\mu_0\to\infty$, we get back the result for resetting with refractory period; (ii) $\gamma\gg\mu_0$ where the particle relaxes to a Boltzmann distribution of an Ornstein-Uhlenbeck process  with half the strength of the original potential and (iii) intermediate $\gamma=2n\mu_0$ for $n=1, 2$. The mean first passage time (MFPT) to find a target  exhibits an optimisation with the switching rate, however unlike instantaneous resetting the MFPT does not diverge but reaches a stationary value at large rates. MFPT also shows similar behavior with respect to the potential strength.  Our results can be verified in experiments on colloids using optical tweezers.
\end{abstract}

\section{Introduction}
Brownian motion is a simple stochastic process that has found a wide range of applications across many disciplines including natural sciences \cite{brown1,brown2,brown3}, ecology \cite{eco}, computer science\cite{comp}, and finance\cite{finance}. The standard Brownian motion is described by the Langevin equation
\bea
\dot{x}=\sqrt{2D}\; \eta(t),
\eea
where $\eta(t)$ is the Gaussian white noise with zero mean and delta-correlated two-point correlation, and $D = k_B T/\nu$ (where $\nu$ is the damping coefficient). Here the variance increases linearly with time ($\sigma^2(t)=2Dt$) and the position distribuion never reaches a stationary state. One of the ways to attain a stationary state is to put the Brownian particle in a confining potential---the most popular one being the Ornstein-Uhlenbeck process where the confining potential is harmonic ($V(x)=\mu x^2/2$) \cite{ou}. This process reaches a steady state and the position distribution is given by the corresponding Boltzmann distribution ($\propto \exp(-V(x)/(k_B T))$).

Another way of reaching a stationary state is by adding a stochastic resetting \cite{evans2011diffusion} to the normal Brownian dynamics. Stochastic resetting refers to random interuptions and restarting of a dynamical process \cite{evans2020stochastic}. Early motivation for stochastic resetting was because of its relevance to search processes where a searcher tries to find a target object for a while, upon an unsuccessful attempt, it returns to the initial location and restarts the process of searching. Since the last decade stochastic resetting has made a profound impact in the field of nonequilibrium statistical physics because of its rich features--- attainment of a nonequilibrium stationary state at long times\cite{evans2011diffusion}, optimisation of search times \cite{evans2011diffusion2}, dynamical transition in relaxation to the stationary states, etc. The effect of resetting has been studied in a wide range of systems---diffusive processes such as
Brownian motion, random walks and L\'evy walks and L\'evy flights\cite{montero2013monotonic, pal2015diffusion, gupta2019stochastic,  shkilev2017continuous, montero2017continuous,kusmierz2014first, kusmierz2015optimal,zhou2020continuous, convexhull}, random acceleration process~\cite{rap}, active particles~\cite{resetrtp1,resetrtp2,resetabp1,resetabp2}, enzymatic reactions~\cite{enzyme1,enzyme2}, active transport in living cells~\cite{activetransport}, fluctuating interfaces~\cite{interface1,interface2}, reaction-diffusion systems~\cite{reactdiff}, Ising model with Glauber
dynamics\cite{isingreset}, asymmetric exclusion processes~\cite{sepreset1,sepreset2}. Most of the works consider instantaneous resetting and restart, however recently the effect of refractory period has been studied~\cite{refractoryperiod} where the particle is reset and remains inactive at the resetting position for sometime before restarting. 

In spite of vast amount of theoretical works, experiments in resetting have been very limited \cite{besga2020optimal,tal2020experimental} due to the challenging nature of the setup. In \cite{besga2020optimal} an optical tweezer is turned on, kept on for a pre-determined period and then turned off. During the time period when the tweezer is turned on no measurements are made. In the other experiment \cite{tal2020experimental} the particle diffuses freely and after exponentially distributed time intervals the particles are driven back to the starting position mimicking resetting events. Similar theoretical models with different return protocols have been proposed recently~\cite{pal19,bodrova2020_1,bodrova2020_2,pal20,gupta2020resetting,gupta2020stochastic}.

In this paper we consider a simple scenario where a particle executing standard Brownian motion is subjected to an intermittent harmonic potential switched on and off at a constant rate. We show that this process retrieves many known results and properties of stochastic resetting in certain regimes. Clearly, switching off the potential allows the free Brownian motion of the particle, while the on-state produces an attractive motion of the particle towards the centre of the potential, which actually brings in the effect of resetting. One advantage of this process is that an experimentalist does not need to track the return or drive the particle to the resetting position in a deterministic way. A variant of this problem was studied very recently where linear confining potential was considered~\cite{mercado2020intermittent}. However, harmonic potentials are much easier to set up in experiments using optical tweezers and any trapping potential can be approximated to a harmonic potential near its minima. From this viewpoint our findings in this paper can easily be verified in experiments.

The rest of the paper is arranged as follows.
First, in \sref{model}, we discuss the model briefly.
Next, in \sref{moments} we study the time-dependent behavior of the variance of the displacement starting with the both off and on state of the potential.
In \sref{stationary_distribution} we discuss the stationary state distribution of the displacement in detail with a brief discussion about the mean first passage time in \sref{mfpt}.
Finally, we summarize our findings with conclusions in \sref{conclusion}.

\section{Model and Results}
\label{model}

\begin{figure}[h]
    \centering
    \includegraphics[scale=0.5]{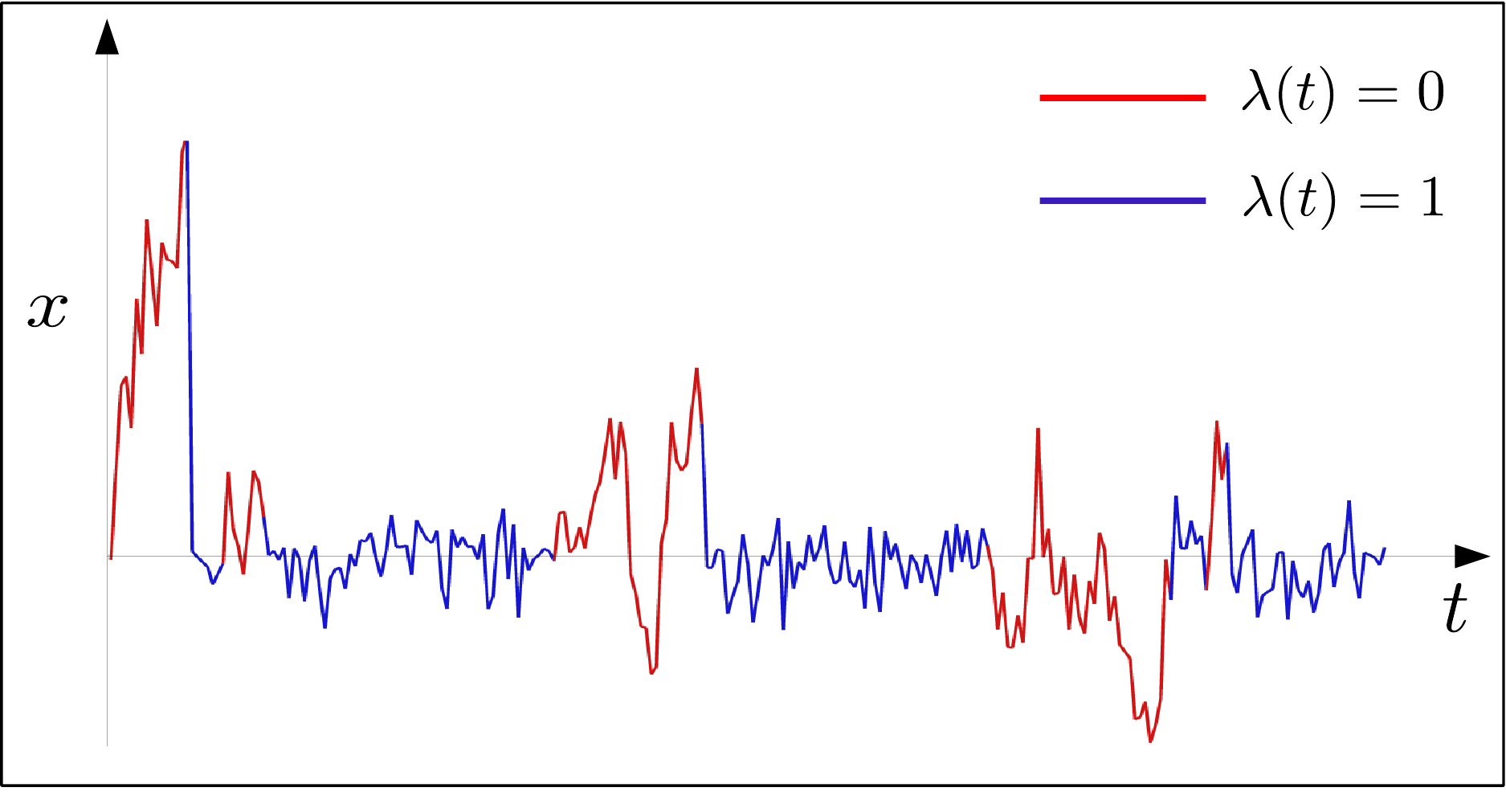}
    \caption{Typical trajectory of a particle under intermittent harmonic trap which switches on and off at rate $\gamma$. The red denotes phases when the harmonic trap is off, while the blue denotes phases when the harmonic trap is on.}
    \label{fig:trajectory}
\end{figure}

We consider a Brownian particle in a stochastically fluctuating, confining potential $V(x,t)$ which is turned on or off stochastically at a constant hazard rate $\gamma$. Mathematically this can be modelled by the Langevin equation
\bea
\dot{x}&=& - \frac{1}{\nu}~\partial_x V(x,t) + \sqrt{2D} \;\eta(t)\qquad \text{with}\qquad V(x,t)=\lambda(t) V(x)
\label{eq:model}
\eea
where $\lambda(t)$ is a dichotomous noise that switches between $0$ and $1$, stochastically, at a constant rate $\gamma$ and $\eta(t)$ is a Gaussian white noise with zero mean and delta-correlator $\langle \eta(t) \eta(t')\rangle = \delta(t -t')$. 
We consider $V(x)$ to be a harmonic potential $\mu x^2/2$. 
Switching off the potential allows the overdamped Brownian motion of the particle of the form $\dot{x}(t) = \sqrt{2D}~ \eta(t)$.
As soon as the harmonic trap is turned on, the dynamics of the particle becomes an Ornstein-Uhlenbeck process $\dot{x}(t)=-\mu_0 x(t) + \sqrt{2D}~\eta(t)$, where $\mu_0 = \mu/\nu$ can be thought of as an effective potential strength, normalized by the damping coefficient. From this point onward, we will refer the potential strength in terms of this normalized strength $\mu_0$. Figure \ref{fig:trajectory} shows a typical trajectory of a particle undergoing the dynamics described by Eq.~\eref{eq:model}.

In the following, we summarize the main results obtained in this paper. 
\begin{itemize}
\item We calculate the full time dependent variance and see that this process always reaches a nonequilibrium stationary state. The leading order relaxation time comes out to be $(\gamma+\mu_0-\sqrt{\gamma^2+\mu_0^2})^{-1}$ beyond which the variance relaxes to the stationary value $D(2/\mu_0+1/\gamma)$.
\item We solve the stationary Fokker-Planck equation and obtain the exact characteristic function. This we invert  in certain limiting cases to obtain the exact stationary state distributions, which show very interesting behavior (see \fref{phase}): (i) For $\gamma\ll\mu_0$ a distinct central Gaussian region followed by exponential tails is seen. The exponential tails have the exact same decay exponent as that of a standard Brownian particle undergoing resetting events at a constant rate, with the switching rate playing the role of the resetting rate. The central Gaussian region becomes narrower as $\mu_0$ keeps on increasing and in the limit $\mu_0\to\infty$ it becomes a $\delta$-function---which is precisely the result for diffusion in presence of stochastic resetting with Poissonian refractory periods. (ii) for $\gamma\gg\mu_0$ the stationary distribution is same as that of an Ornstein-Uhlenbeck process, but with a trap strength $\mu_0/2.$ We also calculate the distribution for some specific intermediate values of $\gamma$ and $\mu_0$ and predict the general functional form for the stationary distribution to be a combination of Gaussian and exponential function.
\begin{figure}[h]
    \centering
    \includegraphics[width=13.cm]{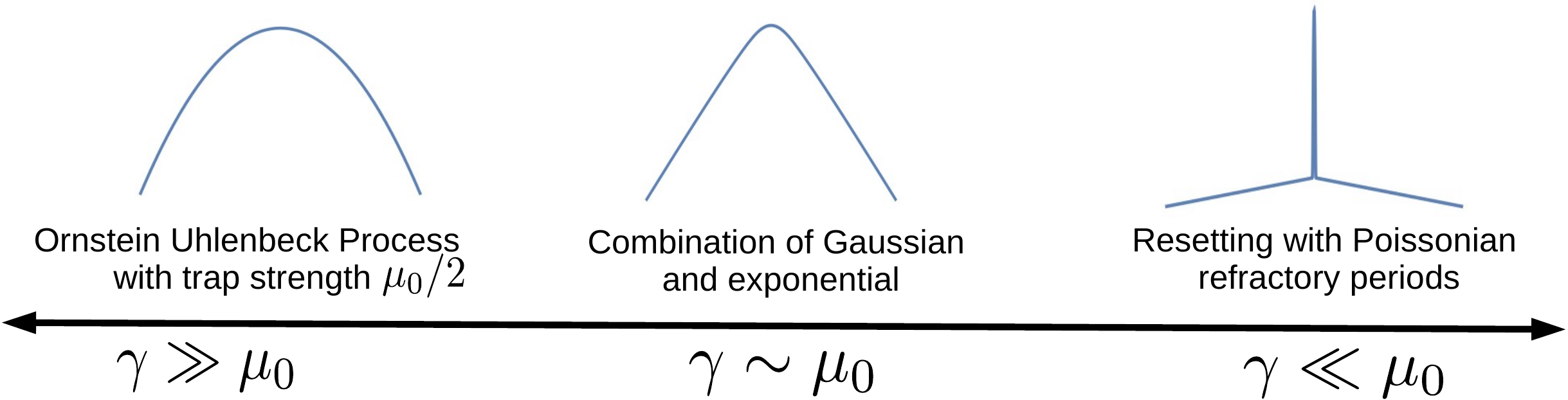}
    \caption{Stationary state phase diagram for a Brownian motion under intermittent harmonic potentials $\mu_0 x^2/2$ showing the different limiting behaviors.}
    \label{phase}
\end{figure}
\item We numerically investigate the mean first-passage time (MFPT) for this process in presence of an absorbing boundary away from the minima of the harmonic potential. For a fixed $\mu_0$ the MFPT reaches a minima for a particular value of $\gamma$, however for very large $\gamma$ it saturates to a constant value unlike in the case of resetting where MFPT actually diverges as the resetting rate goes to $\infty.$    For a fixed $\gamma$ the variation of MFPT with $\mu_0$ shows a similar behavior i.e., it reaches a minima and then saturates to a constant value for large $\mu_0$. We analytically predict the saturation values in both cases which shows excellent agreements with our simulations.
%We correctly predict the saturation value in both cases.
\end{itemize}

In the following section we compute the full time dependent variance for the dynamics.

\section{Moments}
\label{moments}

An analysis of the moments provides a good basic understanding of a stochasitc process. For this process, the Gaussian nature of the white noise and the symmetry of the trap ensures that the distribution is symmetric at all times and thus all the odd moments of the distribution vanish. In this section we look at the time evolution of the first non-zero moment---the variance. Unlike normal Brownian motion, Ornstein-Uhlenbeck process or instantaneous resettings, it is difficult to calculate the exact time dependent moments from the Langevin equations directly, so we use a different procedure. At a particular time $t$ the trap can be on or off, thus the probability that the particle is at position $x$ at time $t$ has two components---$\pon(x,t)$ and $\pof(x,t)$, with the total probability being $P(x,t) = \pon(x,t) + \pof(x,t)$. The full Fokker-Planck equations governing the time evolutions of these probabilities can be written as,
\bea
\hspace{-1cm}\frac{\partial\pon(x,t)}{\partial t}=\mu_0\frac{\partial }{\partial x}(x\pon(x,t)) +D\frac{\partial^2 \pon{(x,t)}}{\partial x^2}  -\gamma \pon(x,t)+\gamma \pof(x,t),\label{eq:fp_xt_on} \\
\hspace{-1cm} \frac{\partial\pof(x,t)}{\partial t}=D\frac{\partial^2 \pof{x,t}}{\partial x^2}-\gamma \pof(x,t)+\gamma \pon(x,t).\label{eq:fp_xt_of}
\eea
First we take the Fourier transform with respect to the position 
variable $x$, and then the Laplace transform with respect to the 
time variable $t$ of both the Eqs. \eref{eq:fp_xt_on} and  
\eref{eq:fp_xt_of},
\begin{eqnarray} 
 \label{fponFLic_m}
s\widetilde{P}_{\text{on}}(k,s) = 
 -Dk^2\widetilde{P}_{\text{on}}(k,s) 
-\mu_0 k\frac{\partial\widetilde{P}_{\text{on}}(k,s)}{\partial k}
-\gamma\widetilde{P}_{\text{on}}(k,s)+\gamma\widetilde{P}_{\text{off}}(k,s).\\
\label{fpoffFLic_m}
s\widetilde{P}_{\text{off}}(k,s)-1 = 
-Dk^2\widetilde{P}_{\text{off}}(k,s) 
-\gamma\widetilde{P}_{\text{off}}(k,s)
+\gamma\widetilde{P}_{\text{on}}(k,s), 
 \end{eqnarray} 
 where we use the conventions for Fourier and Laplace transforms as 
\begin{eqnarray} 
 \label{flCon}
\hspace{-1cm} \widehat{P}_{j}(k,t) = \int_{-\infty}^{\infty}P_j(x,t)~e^{-ikx}~dx, 
 ~~\mathrm{and}~~ 
\widetilde{P}_{j}(k,s) = \int_{0}^{\infty}\widehat{P}_j(k,t)~e^{-st}~dt, 
 \end{eqnarray} 
 respectively for the index $j\in\{\mathrm{on}, \mathrm{off}\}$. We have taken the initial condition that the particle 
 starts from the origin in the off state i.e.,  $P_{\text{off}}(x,t=0)=\delta(x)$, and $P_{\text{on}}(x,t=0)=0$. 
  
\begin{figure}[t]
%\centering
\hspace{2cm}
\includegraphics[scale=0.65]{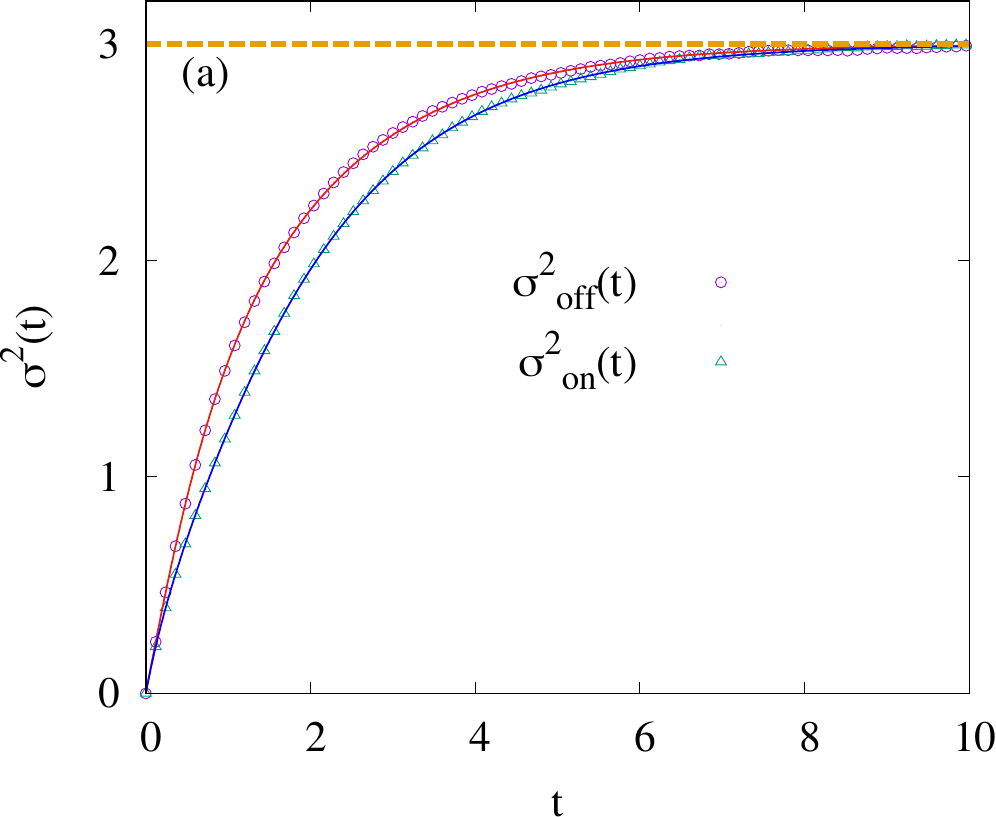}~
\includegraphics[scale=0.335]{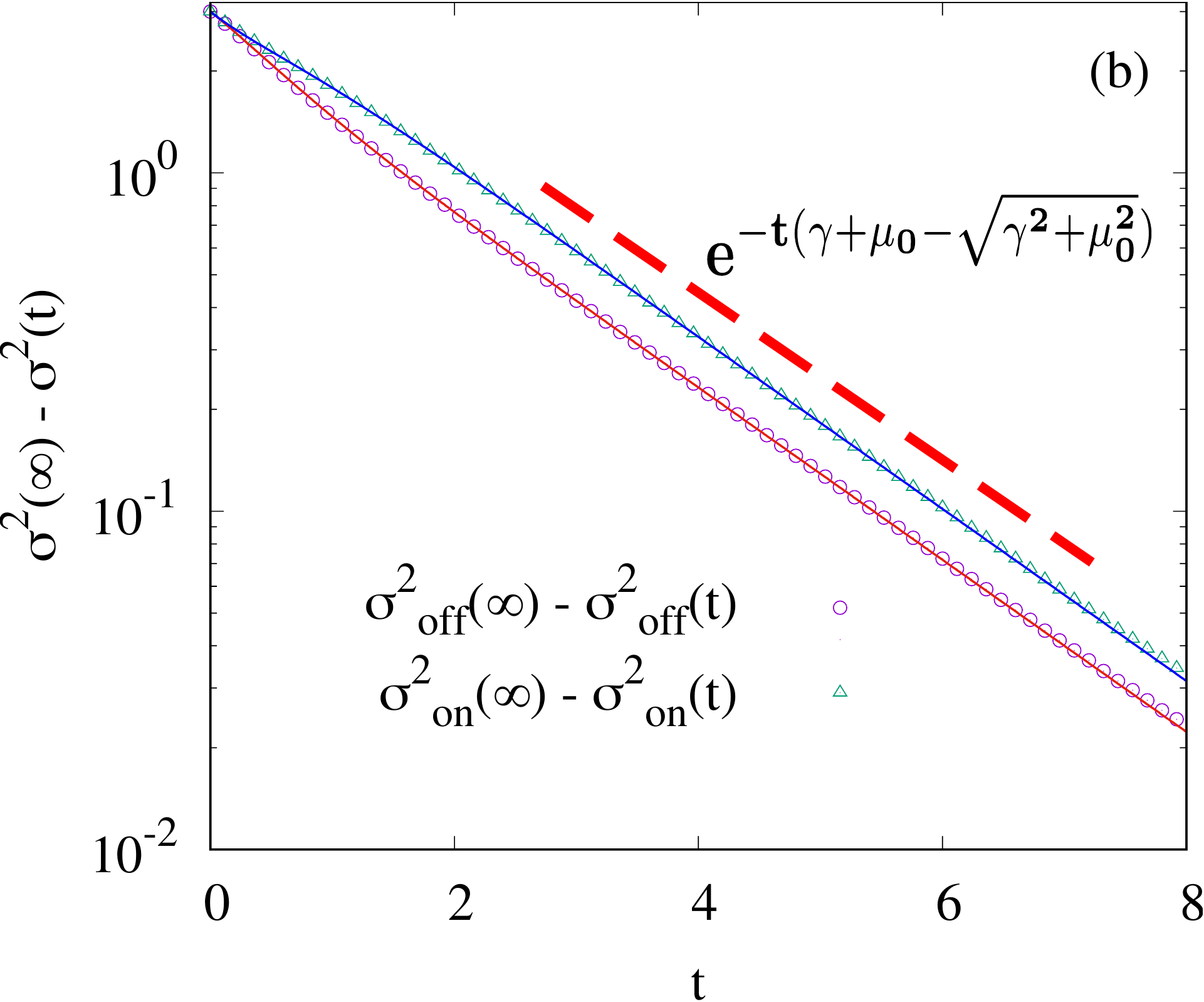}
\caption{ Panel (a): Variance of position of the particle is plotted versus time for the both off and on initial state of the harmonic potential. Discrete symbols are simulation results which are showing excellent agreements with the analytical results of Eqs.~\eref{vvartime} and \eref{vvartimerev} respectively as shown by solid lines. The dashed horizontal line indicates the stationary value of the variance. Panel (b) shows the approach of the time dedpendent variance to its stationary value. Red dashed line denotes the analytical slope of the form of Eq. \eref{sigma_2_slope}. In both plots $D =\mu_0 = \gamma = 1$. Simulation results are averaged over $8 \times 10^5$ number of realizations with time step of integration $\Delta t = 10^{-3}$.}
\label{fig:var_onoff_offon}
\end{figure}

Solving for $\widetilde{P}_{\text{off}}$ from Eq. \eref{fpoffFLic_m} 
as 
%%%%%%%%%%%%%%%%%%%%%%%%%%%%%%%%%%%%%%%%%%%%%%%%%% 
\begin{eqnarray}  
\label{fpoffFLsol1_m}
\widetilde{P}_{\text{off}}(k,s) &=& 
\frac{\gamma}{Dk^2+s+\gamma}\widetilde{P}_{\text{on}}(k,s) 
+\frac{1}{Dk^2+s+\gamma}, 
\end{eqnarray} 
%%%%%%%%%%%%%%%%%%%%%%%%%%%%%%%%%%%%%%%%%%%%%%%%%% 
and substituting its value in Eq. (\ref{fponFLic_m}) we obtain 
%%%%%%%%%%%%%%%%%%%%%%%%%%%%%%%%%%%%%%%%%%%%%%%%%% 
\begin{eqnarray} 
\label{fponFLsol1}
\frac{\partial\widetilde{P}_{\text{on}}(k,s)}{\partial k} &=& f(k,s)
\widetilde{P}_{\text{on}}(k,s) +g(k,s). 
\end{eqnarray} 
%%%%%%%%%%%%%%%%%%%%%%%%%%%%%%%%%%%%%%%%%%%%%%%%%% 
where, $f(k,s) = -\frac{1}{\mu_0 k}
\left(Dk^2+s+\gamma-\frac{\gamma^2}{Dk^2+s+\gamma}\right)$ and $g(k,s)=\frac{\gamma}{\mu_0 k(Dk^2+s+\gamma)}$. The general solution of the above differential equation can be written as,
\begin{eqnarray} 
 \label{fponFLsol2_m}
\hspace{-1cm} \widetilde{P}_{\text{on}}(k,s) = 
 \widetilde{P}_{\text{on}}(0,s)e^{\int_0^kf(k_1,s)dk_1} 
 +e^{\int_0^kf(k_1,s)dk_1}\int_0^kg(k_1,s)e^{-\int_0^{k_1}f(k_2,s)dk_2}, 
 \end{eqnarray} 
%%%%%%%%%%%%%%%%%%%%%%%%%%%%%%%%%%%%%%%%%%%%%%%%%% 
 where $k_1$ and $k_2$ are introduced just as the dummy variables of 
 integrations with respect to the Fourier variable. The total 
 probability density is $P(x,t)=P_{\text{off}}(x,t)+P_{\text{on}}(x,t)$, which implies $\widetilde{P}(k,s)=\widetilde{P}_{\text{off}}(k,s)+\widetilde{P}_{\text{on}}(k,s)$ in the Fourier-Laplace space. Thus, using Eq. (\ref{fpoffFLsol1_m}) we have 
 \begin{eqnarray} 
 \label{ptotFL_m} 
 \nonumber 
\widetilde{P}(k,s)&=&\widetilde{P}_{\text{off}}(k,s)+\widetilde{P}_{\text{on}}(k,s) \\ 
&=& \widetilde{P}_{\text{on}}(k,s)\left(1+\frac{\gamma}{Dk^2+s+\gamma}\right)
+\frac{1}{Dk^2+s+\gamma}. 
 \end{eqnarray} 
 Inverting the expression of $\widetilde{P}(k,s)$ in 
 Eq. \eref{ptotFL_m} to get its value in the $(x,t)$ 
domain is a highly nontrivial task. We, however, are interested in obtaining the time evolution of the moments, for which we need the derivatives of Eq.~\eref{ptotFL_m} as $k\to 0$.
The general relation between the $n^{th}$ moment of a distribution $p(x)$ and its Fourier transform $\hat{p}(k)$ is given by
\bea
\nonumber
\langle x^n \rangle =  (-i)^n\; \displaystyle{\lim_{k \to 0}}\;\frac{\partial^n\hat{p}(k)}{\partial k^n} ,
\eea
when the $n^{th}$ absolute moment of $p(x)$ exists.
Therefore, from Eq.~\eref{ptotFL_m} we calculate the second derivative of $\widetilde{P}(k,s)$ with respect to $k$ as 
%%%%%%%%%%%%%%%%%%%%%%%%%%%%%%%%%%%%%%%%%%%%%%%%%% 
\bea
 \label{der2ptotFL} 
 \nonumber 
\frac{\partial^2\widetilde{P}(k,s)}{\partial k^2} = 
 \gamma\widetilde{P}_{\text{on}}(k,s) \left(\frac{8 D^2 k^2}{\left(\gamma + D k^2+s\right)^3}-\frac{2 D}{\left(\gamma +D k^2+s\right)^2}\right)
\\-\frac{4 \gamma D k }{\left(\gamma +D k^2+s\right)^2}
\frac{\partial\widetilde{P}_{\text{on}}(k,s)}{\partial k}  
\nonumber 
+\frac{\partial^2\widetilde{P}_{\text{on}}(k,s)}{\partial k^2} 
\left(\frac{\gamma }{\gamma +D k^2+s}+1\right)
\\+\frac{8 D^2 k^2}{\left(\gamma +D k^2+s\right)^3}
-\frac{2D}{\left(\gamma +D k^2+s\right)^2}. 
\eea 
The full-time dependent variance in the Laplace domain is obtained by taking the limit $k\to 0$ of the above equation. Taking this limit is quite non-trivial and has been worked out in detail in the \aref{secgentime}. We quote the final result here,
\bea
 \label{vvartime} 
 \nonumber 
 \sigma_{\text{off}}^2(t)=
\frac{D(2\gamma+\mu_0)}{\gamma\mu_0}-\frac{De^{-2\gamma t}}{\gamma}-\frac{D}{\mu_0}\cosh(t\sqrt{\gamma^2+\mu_0^2})\\-\frac{D e^{-t(\gamma+\mu_0)}}{\mu_0\sqrt{\gamma^2+\mu_0^2}}
\left((\gamma+\mu_0)\sinh(t\sqrt(\gamma^2+\mu_0^2))\right),
 \eea
where $\sigma_{\text{off}}^2(t)$ denotes the variance when the particle starts from off state. Similarly, if we start from the on state, the initial conditions used in Eqs. \eref{fponFLic_m} and \eref{fpoffFLic_m} change (see details in the \aref{secgentime}) and the corresponding variance turns out to be
\bea
 \label{vvartimerev} 
 \nonumber 
 \sigma_{\text{on}}^2(t)= 
 \frac{D(2\gamma+\mu_0)}{\gamma\mu_0}+\frac{D e^{-2\gamma t}}{\gamma }-\frac{D\sqrt{\gamma ^2+\mu_0 ^2}}{\gamma \mu_0 }e^{-t(\gamma+\mu_0)} \sinh(t\sqrt{\gamma^2+\mu_0^2})\\-\frac{D(\gamma+\mu_0)}{\gamma\mu_0}e^{-t(\gamma+\mu_0)}\cosh(t\sqrt{\gamma^2+\mu_0^2}).
 \eea
Taking limit $t \rightarrow \infty$ in Eqs.~\eref{vvartime} and \eref{vvartimerev} we find that the two variances tend to the same steady state value $D(2/\mu_0+1/\gamma)$, irrespective of the initial state.
This observation indicates that the particle forgets its initial condition after sufficiently long time.
The predictions in Eqs.~\eref{vvartime} and \eref{vvartimerev} are compared with numerical simulations in Fig. \ref{fig:var_onoff_offon} (a). 
The variance reaches the stationary value exponentially,
\bea
\sigma^2(t\to\infty)-\sigma^2(t)\sim e^{-t\left(\gamma+\mu_0-\sqrt{\gamma^2+\mu_0^2}\right)}.
\label{sigma_2_slope}
\eea
We compare the decay in Eq. \eref{sigma_2_slope} with numerical simulations in Fig \ref{fig:var_onoff_offon} (b).

%This decay is compared with numerical simulations in Fig \ref{fig:var_onoff_offon} (b).

%--------------------------------
%--------------------------------

\section{Stationary Distribution}
\label{stationary_distribution}
Having an indication from the calculation of the moments that the distribution reaches a stationary state, we try to obtain the corresponding nonequilibrium stationary (NESS) distribution.
At large time, we assert that both $P_{\text{on}}$ and $P_{\text{off}}$ become independent of time,  individually.
Therefore, to obtain the stationary distribution of such states we set the lhs of the Fokker-Planck Eqs.~\eref{eq:fp_xt_on}, \eref{eq:fp_xt_of} to $0,$ to obtain 
\bea
\mu_0\frac{\partial }{\partial x}(x\pon(x))+D\frac{\partial^2 \pon (x)}{\partial x^2}-\gamma \pon(x)+\gamma \pof(x)&=&0,\label{eq:fp_ss_on}\\
D\frac{\partial^2 \pof{(x)}}{\partial x^2}-\gamma \pof(x)+\gamma \pon(x)&=&0.\label{eq:fp_ss_of}
\eea
Note that we use the same notation for the stationary probabilities as the time dependent ones for simplicity.
To solve the above equations, it is easier to work in the Fourier space where the equation governing $\pof(x)$ becomes an algebraic one. Thus upon doing a Fourier transform as defined earlier, we have
\bea 
\label{hon_k}
-Dk^2\hon(k)-\mu_0\frac{\partial}{\partial k}(k\hon {(k)})+(\mu_0-\gamma)\hon {(k)} +\gamma\hof {(k)} &=&0,\\
-D k^2\hof(k)-\gamma\hof {(k)} +\gamma\hon {(k)} &=&0.\label{hofk}
\eea
Writing $\hof(k)$ in terms of $\hon(k)$, using Eq.~\eref{hofk}, and replacing it in Eq. \eref{hon_k} we get a single differential equation in terms of $\hon(k)$ as,
\bea
\mu_0 \frac{\partial \hon(k)}{\partial k}+D k\left(1+\frac{ \gamma}{\gamma+Dk^2}\right)\,\hon(k)&=&0.
\label{mu_0_st}
\eea
The solution of Eq. \eref{mu_0_st} can be easily obtained as
\bea 
\hon(k)=\frac{C_0}{(\gamma+Dk^2)^{\gamma/(2\mu_0)}}\, e^{-Dk^2/(2\mu_0)}.
\eea
where $C_0$ is a numerical constant independent of $k.$ Now, $C_0$ can be obtained to be $\gamma^{\gamma/(2\mu_0)}/2$ using the fact that $\hat{P}(0)=\hon(0)+\hof(0)=1.$ Thereafter, using Eq.~\eref{hofk}, we have the full distribution in $k$-space as
\bea
\hspace{-1cm} \hat{P}(k)&=&\frac{e^{-Dk^2/(2\mu_0)}}{2(1+Dk^2/\gamma)^{\gamma/(2\mu_0)}}\left(1+\frac{1}{1+Dk^2/\gamma}\right).\label{eq:pkfull}
\eea
%This is a nonequilibrium stationary state and the factor $1/2$ denotes the contribution to the stationary distribution from the on and off states---this is true for any exponentially switching two-state process.
% One can get the stationary values of all the moments of the distribution by taking derivatives of Eq.~\eref{eq:pkfull}, provided the moments exist. 
To get the stationary distribution in real space we have to invert $\hat{P}(k)$, which unfortunately does not yield any closed form expression for any general values of $\gamma$, $D$, and $\mu_0$. However, it turns out that we can write $P(x)$ as a convolution, which gives us information about some of the asymptotes of the distribution in real space. We can rewrite Eq.~\eref{eq:pkfull} as,
\bea
\hat{P}(k)&=&\frac 1 2\left(\frac{e^{-Dk^2/(2\mu_0)}}{f_1(k)^{\alpha}}+\frac{e^{-Dk^2/(2\mu_0)}}{f_1(k)^{1+\alpha}}\right),
\label{pkfull:eq}
\eea
where $f_1(k)=(1+Dk^2/\gamma)$ and $\alpha=\gamma/{2\mu_0}$. The inverse Fourier transform of the individual terms can be evaluated exactly as,
\bea
\label{g1}
\hspace{-1cm}{\cal F}^{-1}[e^{-Dk^2/(2\mu_0)}]=\frac{1}{\sqrt{2\pi D/\mu_0}}e^{-\frac{\mu_0 x^2}{2D}}=g_1(x)\\
\label{g2}
\hspace{-1cm} {\cal F}^{-1}\left[f_1(k)^{-\alpha}\right]=\frac{\sqrt{\pi}}{\Gamma (\alpha)}
\; \left( \sqrt{\frac{\gamma }{D}} \frac{\left| x\right|}{2}\right)^{\alpha-1/2} \; K_{\frac{1}{2}- \alpha}\left( \sqrt{\frac{\gamma }{D}}\left| x\right|\right) =g_2(x,\alpha)
\eea
where $K_n(z)$ is the modified Bessel function of the second kind. Thus, the full distribution is,
\bea
\hspace{-1cm} P(x)&=&\frac{1}{2}\left(\int_{-\infty}^{\infty}dy\, g_2(y,\alpha)g_1(x-y)+\int_{-\infty}^{\infty}dy\, g_2(y,\alpha+1)g_1(x-y) \right).
\label{px_full}
\eea
This integral gives the exact stationary state of the particle for any value of $\gamma,$ $D$ and $\mu_0$. Unfortunately, a closed form expression for the above integral is difficult to obtain. However, for asymptotic parameter values we do obtain the limiting distributions exactly.

If the rate of switching the trap is very small with respect to the strength of the potential, i.e., $\gamma/\mu_0$ is very small, approximating $f_1(k)^{\alpha}$ by its limiting value 1, we obtain
\bea
\hat{P}(k)&\approx&\frac 1 2\left(e^{-Dk^2/(2\mu_0)}+\frac{e^{-Dk^2/(2\mu_0)}}{(1+Dk^2/\gamma)}\right).
\eea
Upon Fourier inversion, the first term yields a Gaussian distribution with variance $D/(2\mu_0)$, while the second term can be evaluated by convolution to obtain,
\bea
\hspace{-2.cm} P(x)&=& \frac12\left(\frac{e^{-\mu_0 x^2/(2D)}}{\sqrt{2\pi D/\mu_0}}+\frac{\sqrt{\gamma/D}e^{\gamma/(2\mu_0)}}{4}e^{-\sqrt{\gamma/D}|x|}\;\text{Erfc}\left(\frac{\sqrt{\gamma/D}}{2\mu_0}- \sqrt{\frac{\mu_0}{2D}}|x|\right)\right),
\label{smallrate}
\eea
where Erfc$(z)$ is the complementary error function. Near the origin the first term dominates, while the behavior at the tails is dictated by the second term.
%---------------------------- 
\begin{figure}[t]
%    \centering
    \hspace{2cm}
    \includegraphics[scale=0.65]{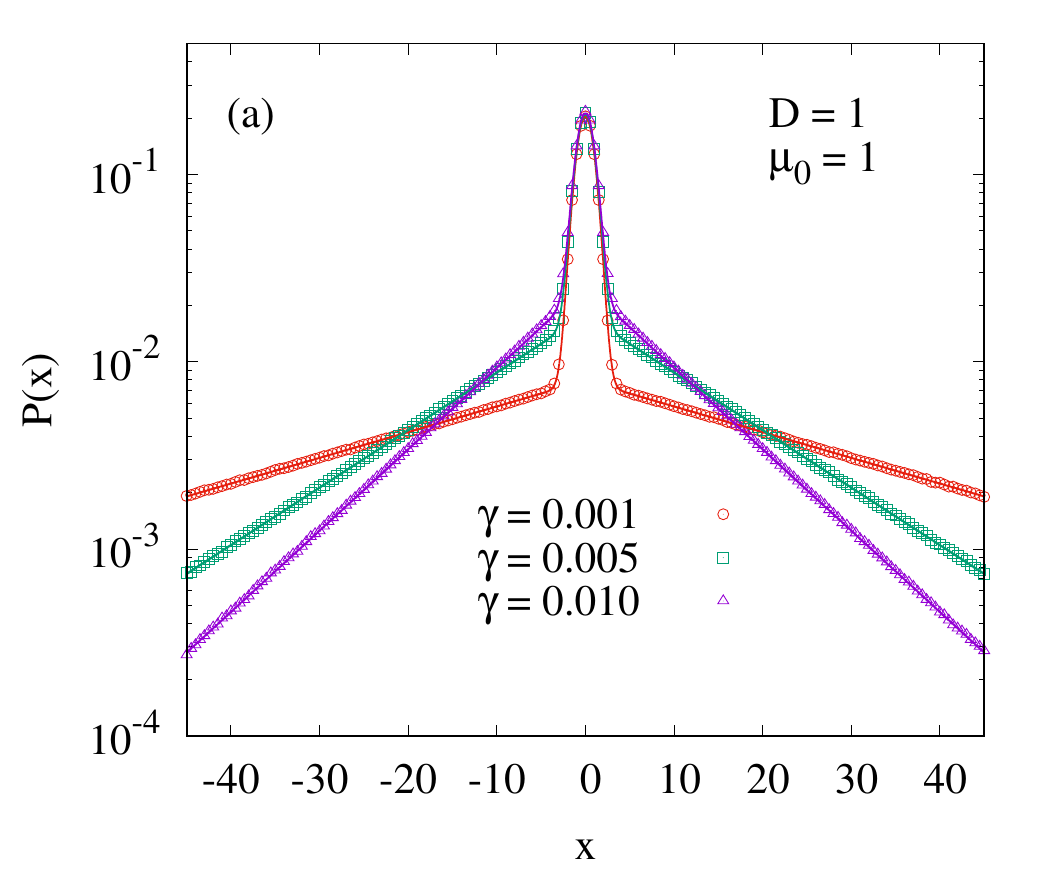}~
    \includegraphics[scale=0.65]{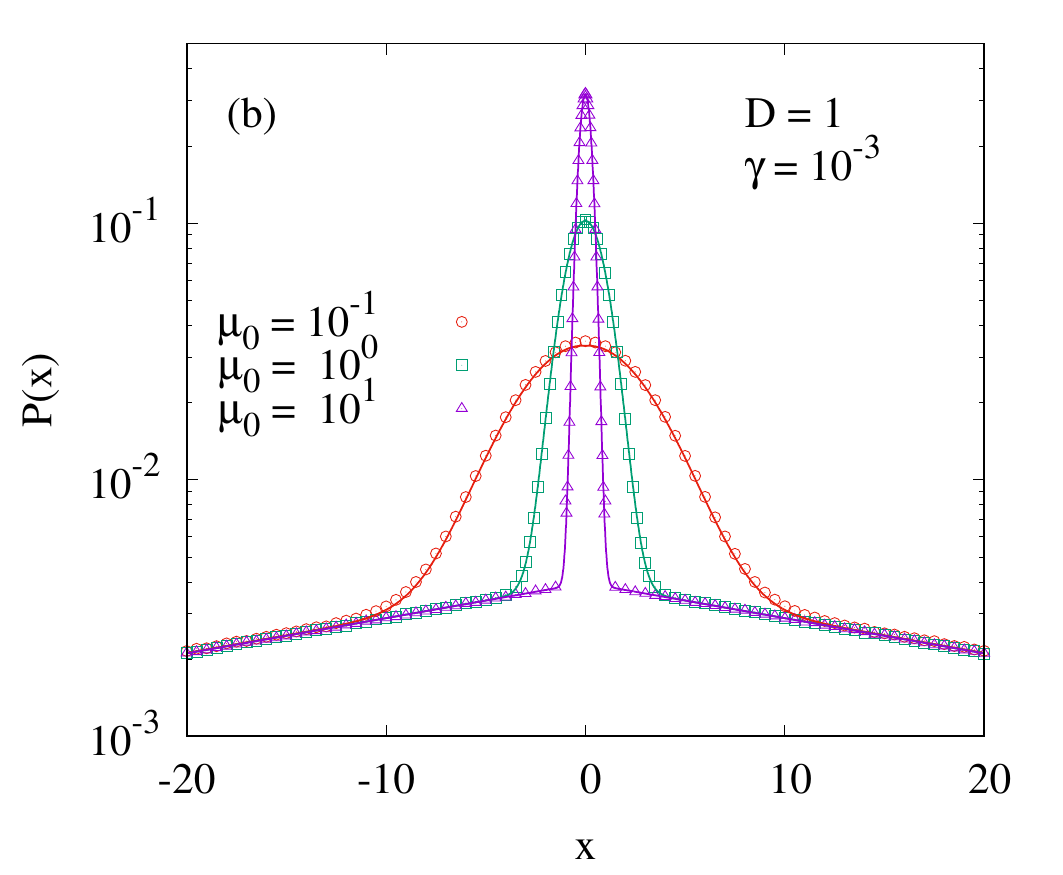}
    \caption{Stationary distribution of position $P(x)$ is plotted in the limit $\gamma \ll \mu_0$ with the diffusion constant $D = 1$. In (a) we plot $P(x)$ for different switching rate $\gamma$ with a fixed $\mu_0 = 1$, whereas in (b) $P(x)$ is plotted for different $\mu_0$ with a fixed $\gamma = 10^{-3}$. Discrete symbols in each plot are from the numerical simulation which are showing excellent agreements with the analytical result of Eq.\eref{smallrate} as shown by solid lines. Simulation results are averaged over $1.28 \times 10^8$ number of realizations with $\Delta t = 10^{-3}$. The stationary state results are obtained by running the simulation for time $t = 10^4$.}
    \label{fig:dist_short_gamma}
\end{figure}
%----------------------------
When the trap is turned on, the typical time taken by a particle to relax to the stationary distribution is $\sim \mu_0^{-1}$, however, here  $\gamma/\mu_0\ll 1$, i.e., the average time to turn the trap off again ($\sim \gamma^{-1}$) is much larger than $\mu_0^{-1}.$ As a result the particle spends some time near the minima of the trap, with the usual Boltzmann distribution (Gaussian distribution for our case) as dictated by the trap, while outside this region the distribution is governed by same exponential tails as the NESS of a diffusion in the presence of instantaneous resetting. The results in Eq.~\eref{smallrate} is compared with the numerical simulations in Fig.~\ref{fig:dist_short_gamma} (a), where we see excellent match. Note that the central Gaussian part becomes narrower as we increase the strength of the potential (as shown in Fig.~\ref{fig:dist_short_gamma} (b)) and in the limit $\mu_0\to\infty$, the first term in Eq.~\eref{smallrate} becomes a $\delta$-function and we get
\bea
P(x)&\approx&\frac 12\left(\delta(x)+\frac{\sqrt{\gamma/D}}{2}e^{-\sqrt{\gamma/D}|x|}\right),
\label{eq:refract}
\eea
where we have also used the limiting value of Erfc$(-z)$ as $z\to\infty$.
When $\mu_0 \to \infty$ the particle returns to the origin 
almost instantaneously as the trap is turned on, and remains there until it is turned off.
Physically, this corresponds to the instantaneous resetting with refractory periods where instantaneous resetting events are followed by a period of immobility at the resetting position.
Indeed Eq.~\eref{eq:refract} is exactly the same as obtained in the reference \cite{refractoryperiod} for resetting with Poissonian refractory periods. In fact, if the switching on and off rates of the trap are considered to be different ($\gamma_{\text{on}}$ and $\gamma_{\text{off}}$, respectively), then one can obtain the results of instantaneous resetting without refractory period~\cite{evans2011diffusion} in the limit $\gamma_{\text{off}}/\gamma_{\text{on}}\ll 1$ (see \aref{different_rates} for a detailed discussion).

\begin{figure}[t]
%    \centering
    \hspace{2cm}
    \includegraphics[scale=0.65]{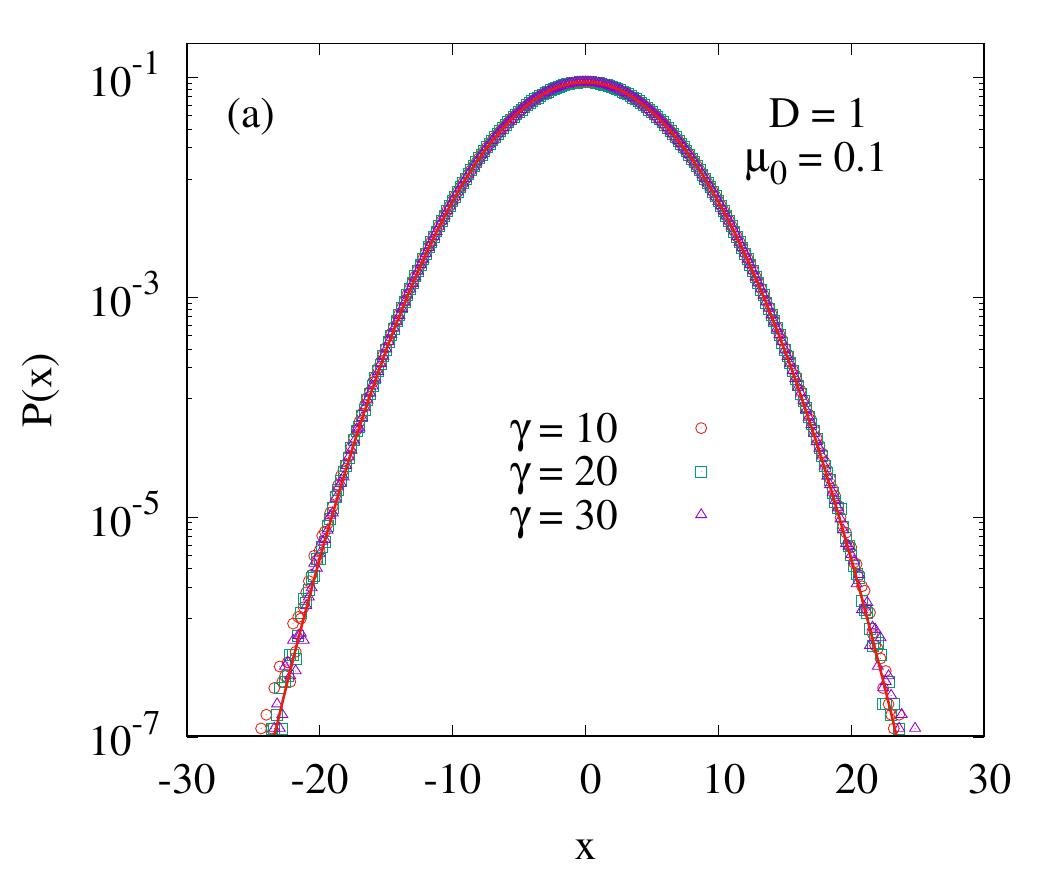}~
    \includegraphics[scale=0.65]{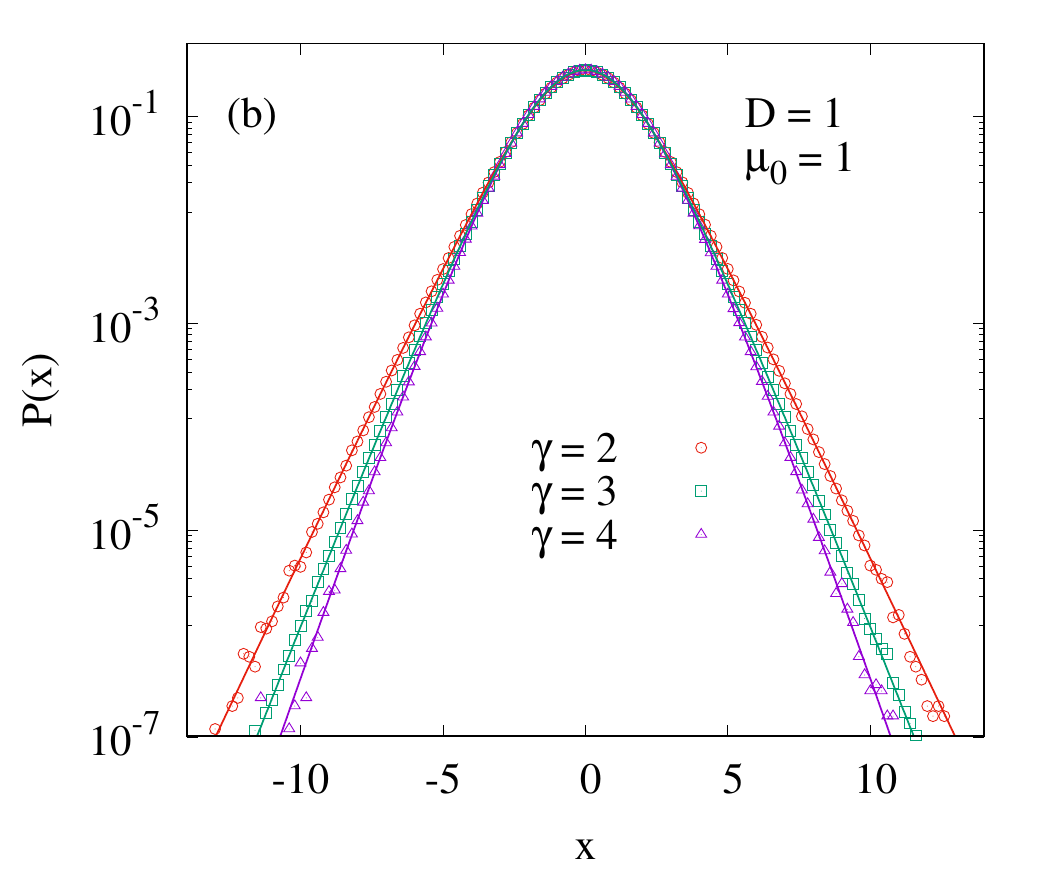}
    \caption{Stationary distribution of position $P(x)$ are plotted in the limits (a) $\gamma \gg \mu_0$ 
    and (b) $\gamma \approx \mu_0$ respectively with the diffusion constant $D = 1$. In both diagrams we plot for different switching rates $\gamma$ with fixed  $\mu_0 = 0.1$ and $\mu_0 = 1$ respectively. Discrete symbols in both plots indicate simulation results which show excellent agreement with the analytical predictions denoted by solid lines. In plot (a) the solid line represents Eq.~\eref{gaussian}; in (b) the solid lines for $\gamma=2,4$ represent Eq.~\eref{gam2},~\eref{gam4} respectively, while the same for $\gamma=3$ is obtained from numerical integration of Eq.~\eref{px_full}). Simulation results are averaged over $1.28 \times 10^8$ number of realizations with $\Delta t = 10^{-3}$. The stationary state results are obtained by running the simulation for time $t = 10^2$.  }
    \label{fig:gam_large_sim}
\end{figure}

On the other hand, when the switching rate of the potential is large with respect to the potential strength $(\gamma/\mu_0 \gg 1)$, Eq.~\eref{eq:pkfull} can be approximated as
\bea
\hat{P}(k) &\approx & e^{-Dk^2/\mu_0},\label{eq;pk_largeg}
\eea
using $(1+Dk^2/\gamma)^{\gamma/(2\mu_0)}\approx(1+Dk^2/\gamma)^{1+\gamma/(2\mu_0)}\approx e^{Dk^2/(2\mu_0)}$ for large $\gamma$. Thus for very large switching rate ($\gamma\to\infty$) the stationary distribution becomes independent of $\gamma$, and Eq.\eref{eq;pk_largeg} upon Fourier inversion, yields a Gaussian distribution
\bea
P(x)\approx \frac{e^{-\mu_0 x^2/4D}}{\sqrt{4\pi D/\mu_0}}.
\label{gaussian}
\eea
Note that this is the Boltzmann distribution of an Ornstein-Uhlenbeck process where the trap strength is $\mu_0/2$. This indicates that as the trap switches on and off very fast (as $\gamma\to\infty$), the particle experiences an average potential of strength $\mu_0/2$ and relaxes to the corresponding Boltzmann distribution. We compare this with numerical simulations in Fig.~\ref{fig:gam_large_sim} (a), and the excellent match confirms our prediction.

%-------------------------------
%--------------------------------

For $\gamma\sim\mu_0 $ it is very difficult to obtain any closed form expression for the stationary state distribution, but one can obtain closed form expressions for $\gamma=2 n \mu_0$, for any fixed integral value of $n$. However, it is difficult to write any closed form expression in terms of an arbitrary integer $n$.
% for $n\in \mathbb{Z}$ using Eq.~\eref{px_full}. 
For example, when $n=1$, we have
\bea
\hspace{-1.5cm} P(x)&=&\frac{e^{-\frac{\mu_0  x^2}{2 D}}}{2 \sqrt{\frac{2 \pi  D}{\mu_0 }}}+\frac{e}{16 \sqrt{\frac{D}{\mu_0 }}} e^{-|x| \sqrt{\frac{2 \mu_0 }{D}}} \left[ \left(2 |x| \sqrt{\frac{\mu_0 }{D}}+\sqrt{2}\right) \text{Erfc}\left(1-|x| \sqrt{\frac{\mu_0 }{2 D}}\right)\right.\cr&+&\left.\left(2 |x| \sqrt{\frac{\mu_0 }{D}}-\sqrt{2}\right) \text{Erfc}\left(1+|x| \sqrt{\frac{\mu_0 }{2 D}}\right)\right].
\label{gam2}
\eea
%\bea 
%\nonumber
%\hspace{-2.8cm} P(x) = \frac{1}{16 D} \sqrt{\frac{\mu_0}{\pi D}}\;\exp\left[{-\frac{\mu_0  x \left(x \sqrt{D \mu_0 }+3 \sqrt{2} D\right)}{\sqrt{D^3 \mu_0 }}}\right] \left(   4 \sqrt{2}\; D \exp \left[{\frac{x \left(6 \sqrt{2} \sqrt{D \mu_0 }+\mu_0  x\right)}{2 D}}\right]  \right. \\
%\nonumber \hspace{-1.5cm}  \left. + \sqrt{\pi }  \;\exp\left[{\frac{2 \sqrt{2} x \sqrt{D \mu_0 }+ D + \mu_0 x^2}{D}}\right]  \left( \left( \sqrt{2} D + 2 x \sqrt{D \mu_0 }\right) \text{erfc}\left( 1 - x \sqrt{\frac{\mu_0}{2 D}}  \right) \right. \right. \\ \left.\left.+ \exp\left[{\frac{2 \sqrt{2} \mu_0  x}{\sqrt{D \mu_0 }}}\right] \left(\sqrt{2} D-2 x \sqrt{D \mu_0 }\right) \text{erfc}\left(1 + x \sqrt{\frac{\mu_0}{2 D}} \right) \right) \right)
%\eea
Similarly, for $n=2$, we have
\bea
\hspace{-2.5cm} P(x)=\frac{3 e^{-\frac{\mu_0  x^2}{2 D}}}{4 \sqrt{2 \pi } \sqrt{\frac{D}{\mu_0 }}}+\frac{e^{- |x| \sqrt{\frac{4 \mu_0 }{D}}}}{32D^{3/2}}e^2 \sqrt{\mu_0 } \left[\left( \left(2 |x| \left(\sqrt{D \mu_0 }+2 \mu_0  |x|\right)-D\right)  \text{Erfc}\left(\frac{|x| \sqrt{D \mu_0 }+2 D}{\sqrt{2} D}\right)\right)\right.\cr
- \left.  \left( \left(2 |x| \left(\sqrt{D \mu_0 }-2 \mu_0  |x|\right)+D\right) \text{Erfc}\left(\frac{|x| \sqrt{D \mu_0 }-2 D}{\sqrt{2} D}\right)\right)\right].
\label{gam4}
\eea
%\bea 
%\nonumber
%\hspace{-2.6cm} P(x) = \frac{1}{32 D} \sqrt{\frac{\mu_0}{\pi D}}\; \exp\left[{-\frac{x \left(4 \sqrt{D \mu_0 }+\mu_0 x\right)}{D}}\right]
%\left(  12 \sqrt{2}\; D \exp\left[{\frac{x \left(8 \sqrt{D \mu_0 }+\mu_0  x\right)}{2 D}}\right] +  \right. \\ \left.\nonumber \sqrt{\pi} \;\exp\left[{\frac{2 x \sqrt{D \mu_0 }+2 D + \mu_0  x^2}{D}}\right] 
%\left( -\left(2 x \left(\sqrt{D \mu_0 }-2 \mu_0  x\right)+D\right)  \right. \right.\\ \left. \left. \nonumber \times \; \text{erfc}\left(\frac{2 D-x \sqrt{D \mu_0 }}{\sqrt{2} D}\right) + \left(2 x \left(\sqrt{D \mu_0 }+2 \mu_0  x\right)-D \right) \exp\left[ \frac{4 \mu_0  x}{\sqrt{D \mu_0 }}\right] \right. \right.\\ \left. \left. \times \; \text{erfc}\left(\frac{x \sqrt{D \mu_0 }+2 D}{\sqrt{2} D}\right) \right) \right).
%\eea
We compare these with the numerical simulations in Fig.~\ref{fig:gam_large_sim} (b), along with the case $\gamma=3\mu_0$. Although there are no formal expressions, since the stationary distribution is a well behaved function for all values of $\gamma,\,\mu_0$ and $D$, as can be understood from its characteristic function, one can anticipate the general form for the stationary distribution to be 
%the general form for the stationary distribution is always of the form
\bea
P(x)\sim e^{-\mu_0 x^2/(2D)}h_1(x^2)+e^{-\sqrt{\gamma/D}|x|}h_2(|x|),
\label{Px_genn}
\eea
where $h_1(x)$ and $h_2(x)$ are polynomials of $x$. Near the origin, the fluctuations are Gaussian, while tails decay exponentially.

\section{Mean First Passage Times}
\label{mfpt}

Another important physical quantity for stochastic problems is the first-passage time distribution. The corresponding first-passage probability $F(x_0,t)dt$ denotes the probability that a particle starting from $x_0$ at $t=0$ reaches the target position (or an absorbing boundary) $x_{abs}$ for the first time between times $t$ and $t+dt$. The mean of this distribution is called the mean first passage time(MFPT). This is particularly relevant in resetting problems as introduction of resetting dynamics optimise the MFPT--- i.e., the mean time to reach the target is minimised for some optimal value of the resetting rate. Here, we have two parameters, namely the potential switching rate $\gamma$ and the trap strength $\mu_0$, we numerically look at how the MFPT depends on these parameters.  

\begin{figure}[t]
%    \centering
    \hspace{2cm}
    \includegraphics[scale=0.68]{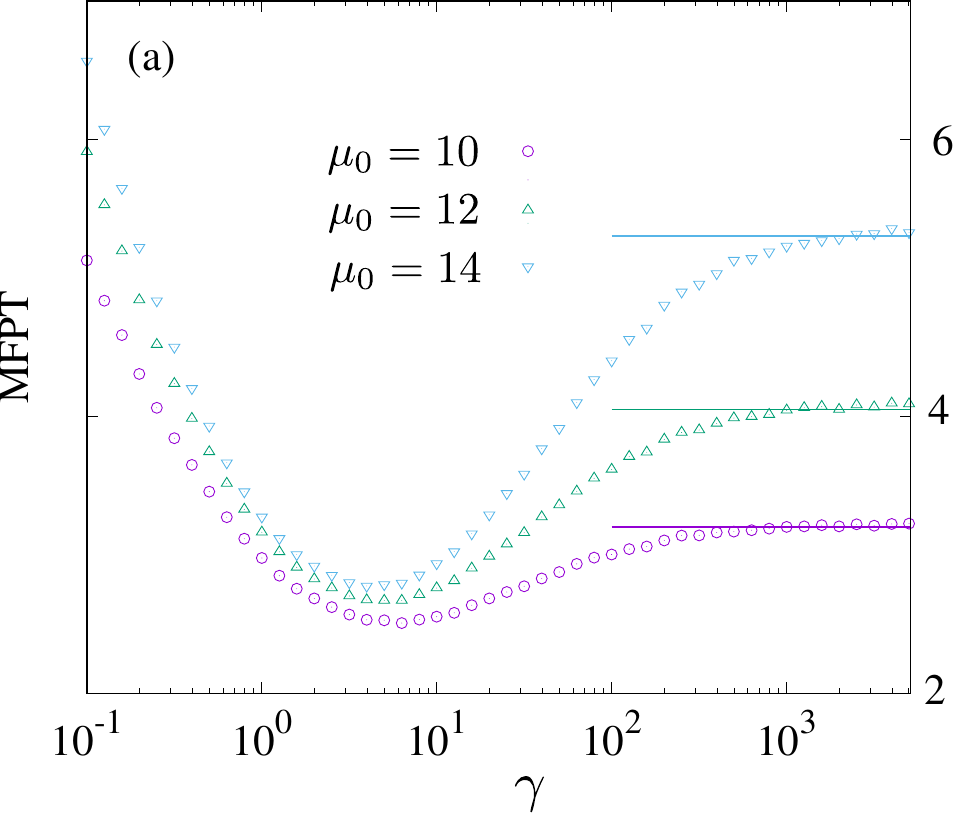}~
    \includegraphics[scale=0.68]{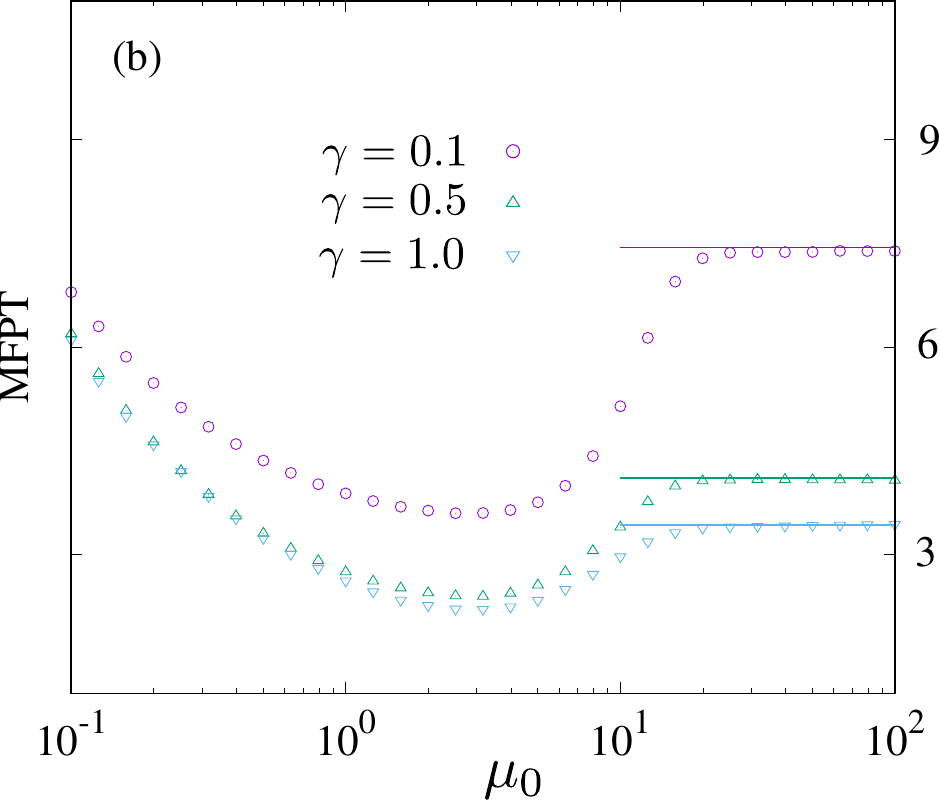}
    \caption{Numerical result of the mean first passage time of a particle  starting from $\lambda(0) = 0$ state at $x(0) = 0$ with an absorbing boundary at $x_{\text{abs}} = -1$ and diffusion constant $D = 1$ are plotted versus the switching rate $\gamma$ and harmonic potential strength $\mu_0$ in (a) and (b) for different $\mu_0$ and $\gamma$ respectively. Solid lines in (a) and (b) correspond to the analytical estimations of the saturation values of MFPT from Eqs. \eref{MFPT_ou_m} and \eref{MFPT_rte} respectively. Simulation results are averaged over $2.56 \times 10^4$ number of realizations with $\Delta t = 10^{-5}$.}
    \label{fig:mfpt}
\end{figure}

We consider a particle starting from $\lambda(t=0)=0$ state from $x(0)=0$ with an absorbing boundary located at $x_{abs}=-1$. Figure \ref{fig:mfpt}(a) shows that for a fixed $\mu_0$, the MFPT shows a non-monotonic behavior--- initially it decreases with increase in $\gamma$, reaches a minimum at some optimal switching rate $\gamma^*$ and then increases with increase in $\gamma$.  When $\gamma\to0$ the dynamics is very much like the normal Brownian motion for which MFPT diverges owing to the trajectories that take particle far away from the target. As $\gamma$ increases the particle switches between a free Brownian particle and an Ornstein Uhlenbeck particle, the effect of the trap forces the particle to return close to the minima of the potential, thus cutting out the trajectories which take larger excursions away from the target. When $\gamma$ is increased further, the MFPT increases and saturates to a constant value. This is unlike instantaneous resetting where it always diverges with increasing resetting rate beyond $\gamma^*$, because in the limit of a very high switching rate, the particle is essentially trapped at the resetting position all the time restricting it from reaching the absorbing boundary in a finite time. However, in our case, we obtain a saturation in MFPT with increasing switching rate. A heuristic argument for this saturation can be given in terms of the stationary distribution in the regime $\gamma\gg\mu_0$ without any absorbing boundaries. The particle actually feels that it is in a harmonic potential of strength $\mu_0/2$ and relaxes to the corresponding Boltzmann distribution which is independent of $\gamma$ as shown in Eq. \eref{gaussian}---thus heuristically, when $\gamma\gg\mu_0$, MFPT saturates to the corresponding stationary value. In fact this stationary value is exactly same as that of an Ornstein-Uhlenbeck process with the trap strength $\mu_0/2$ (keeping the starting position and absorbing boundary same in both cases). The solid lines in Fig.~\ref{fig:mfpt} (a) indicate the values of MFPT for an Ornstein-Uhlenbeck process with trap strength $\mu_0/2$
\bea
\text{MFPT}_{ou}=\lim_{s\to 0}\frac{1}{s} \left( 1-e^{-x_{abs}^2\mu_0/(8D)}\frac{\cal{D}_{-2s/\mu_0}(0)}{\cal{D}_{-2s/\mu_0}(-x_{abs})} \right),
\label{MFPT_ou_m}
\eea
 where $\cal{D}_n$ denotes parabolic cylinder function of order n (See \aref{MFPT_ou}).

If we change the strength of the harmonic potential ($\mu_0$), keeping $\gamma $ fixed then also the MFPT shows a non-monotonic behavior, see Fig.~\ref{fig:mfpt} (b).
For $\mu_0 \to 0$ we recover free Brownian motion where MFPT diverges.
As $\mu_0$ is increased, the MFPT decreases, hits a minima and  eventually reaches a saturation value. 
%In particular, in the limit $\mu_0 \to \infty$ the process becomes equivalent to the instantaneous resetting with refractory period when the rate of the refractory period and the resetting rate are equal.
In this limit, the saturatation of MFPT can be understood heuristically from the distribution without any absorbing boundary for very large $\mu_0$ (which corresponds to the  stochastic resetting with Poissonian refractory periods as shown in Eq.~\eref{eq:refract}), which is independent of $\mu_0$. In fact, the saturation value can be predicted exactly using the result for resetting with refractory periods~\cite{refractoryperiod}
\bea
\text{MFPT}_{\text{ref}}&=&\frac{2}{\gamma}\left(e^{\sqrt{\gamma/D}\,|x_{\text{abs}}|}-1\right)
\label{MFPT_rte}
\eea
(see \aref{mfpt_ref} for more details).
This is plotted using solid lines Fig.~\ref{fig:mfpt} (b) and excellent match confirms our prediction.

\section{Summary and Conclusion}
\label{conclusion}
In this paper we study a Brownian particle under the effect of an intermittent harmonic potential $\mu_0$, which switches on and off at a constant rate $\gamma$. We show that this process reaches a stationary state and calculate the full time dependent variance that gives us the leading order relaxation time scale for the system as $(\mu_0+\gamma-\sqrt{\mu_0^2+\gamma^2})^{-1}$. We then solve the stationary Fokker Planck equation and find the exact characteristic function of the nonequilibrium stationary state. We invert exactly for few special cases: (i) for $\gamma\ll\mu_0$ there is a distinct central Gaussian region followed by exponential tails. The central Gaussian region becomes narrower as $\mu_0$ keeps on increasing and in the limit $\mu_0\to\infty$ it becomes a $\delta$-function---which is precisely the result for diffusion in presence of stochastic resetting with exponentially distributed refractory period. (ii) for $\gamma\gg\mu_0$ the stationary distribution is same as that of an Ornstein-Uhlenbeck process with trap strength $\mu_0/2.$ We also compute the distribution for a few special intermediate cases $\gamma=2\mu_0,\,4\mu_0$ and conclude that for $\gamma \sim \mu_0$ the stationary state is a combination of Gaussian and exponential distributions. 
Finally, we investigate the mean first-passage time numerically---we see that the MFPT is optimised w.r.t. both the switching rate and trap strength when the other is fixed.   
We also numerically investigate the saturation of MFPT with respect to the switching rate and potential strength, and present an interesting heuristic analytical estimation of these saturation values.

%in the limits $\gamma/\mu_0 \rightarrow 0$, $\gamma \sim \mu_0$, $\gamma/\mu_0 \rightarrow \infty$ for 

There are several possible extensions and open questions related to our work. Let us address the theoretical questions first. An obvious question is how the obtained physical behaviors change when we have a general confining potential of the form of $|x|^p$. We look at the MFPT numerically, however it would be interesting to see if one can solve the Fokker-Planck equation with the aborbing boundary conditions exactly and see transitions similar to \cite{kusmierz2014first}. Another generalisation would be to consider the on and off switching rates to be different ($\gamma_{\text{on}}$ and $\gamma_{\text{off}}$ say) as in~\cite{mercado2020intermittent} and see if the MFPT shows transitions in the $\gamma_{\text{on}}-\gamma_{\text{off}}$ plane. One can also apply this protocol on other diffusive models like Levy flights, random acceleration processes, active particles like RTP~\cite{rtp1,rtp2}, ABP~\cite{abp1,abp2} and DRABP~\cite{drabp}.
We expect similar kind of phase diagram (\fref{phase}) for any stochastic process under intermittent attractive potential---at large switching rates the distribution relaxes to the normal steady state distribution, as in the presence of a trap with renormalised trap strength, and to resetting with poissonian refractory periods at very small switching rates. 
Our predictions can be verified in colloidal systems using optical tweezers, for the confining potential in optical traps is inherently harmonic in nature \cite{ashkin1997optical,datar2015dynamics}
, and hence, does not require any additional experimental modifications. 
In fact, experiments following this protocol can also be performed on active matter like bacteria and Janus swimmers. 

\section{Acknowledgements}
The authors thank Urna Basu and Pramod Pullarkat for useful discussions.

%------------------------------------------------
\appendix
\label{appendix}

\section{Fluctuating trap: calculation of relaxation time scales from moments} 
\label{secgentime}

The Fokker-Planck equation for the density function of the position 
of a Brownian particle in a fluctuating harmonic potential can be 
written as 
%%%%%%%%%%%%%%%%%%%%%%%%%%%%%%%%%%%%%%%%%%%%%%%%% 
 \begin{eqnarray} 
 \label{fpoff}
 &&\frac{\partial P_{\text{off}}}{\partial t} = D\frac{\partial^2P_{\text{off}}}
 {\partial x^2}-\gamma P_{\text{off}}+\gamma P_{\text{on}}, \\ 
 \label{fpon}
 &&\frac{\partial P_{\text{on}}}{\partial t} = D\frac{\partial^2P_{\text{on}}}
 {\partial x^2}+\mu_0\frac{\partial}{\partial x}(xP_{\text{on}})-\gamma P_{\text{on}}
 +\gamma P_{\text{off}}. 
 \end{eqnarray} 
 %%%%%%%%%%%%%
We now intend to calculate the time-dependent second moment to 
 understand the relaxation time scales for this system. Therefore, 
 first we take the Fourier transform with respect to the position 
 variable $x$, and then the Laplace transform with respect to the 
 time variable $t$ of both the Equation (\ref{fpoff}) and Equation 
 (\ref{fpon}). This gives us the Equation (\ref{fpoff}) and Equation 
 (\ref{fpon}) in the Fourier-Laplace domain as 
 %
 %%%%%%%%%%%%%%%%%%%%%%%%%%%%%%%%%%%%%%%%%%%%%%%%% 
 \begin{eqnarray} 
 \label{fpoffFL}
\hspace{-2.5cm} s\widetilde{P}_{\text{off}}(k,s)-\widehat{P}_{\text{off}}(k,0) = 
-Dk^2\widetilde{P}_{\text{off}}(k,s) 
-\gamma\widetilde{P}_{\text{off}}(k,s)
+\gamma\widetilde{P}_{\text{on}}(k,s), \\ 
 \label{fponFL}
\hspace{-2.5cm} s\widetilde{P}_{\text{on}}(k,s)-\widehat{P}_{\text{on}}(k,0) = 
 -Dk^2\widetilde{P}_{\text{on}}(k,s) 
-\mu_0 k\frac{\partial\widetilde{P}_{\text{on}}(k,s)}{\partial k}
-\gamma\widetilde{P}_{\text{on}}(k,s)+\gamma\widetilde{P}_{\text{off}}(k,s), 
 \end{eqnarray} 
 %%%%%%%%%%%%%%%%%%%%%%%%%%%%%%%%%%%%%%%%%%%%%%%%% 
 %
 where we use the conventions for Fourier and Laplace transforms as 
 %
 %%%%%%%%%%%%%%%%%%%%%%%%%%%%%%%%%%%%%%%%%%%%%%%%% 
 \begin{eqnarray} 
 \label{flCon}
\hspace{-1cm} \widehat{P}_{j}(k,t) = \int_{-\infty}^{\infty}P_j(x,t)~e^{-ikx}~dx, 
 ~~\mathrm{and}~~ 
\widetilde{P}_{j}(k,s) = \int_{0}^{\infty}\widehat{P}_j(k,t)~e^{-st}~dt, 
 \end{eqnarray} 
 %%%%%%%%%%%%%%%%%%%%%%%%%%%%%%%%%%%%%%%%%%%%%%%%% 
 %
 respectively for the index $j\in\{\text{on}, \text{off}\}$. Assuming that the particle 
 starts from the origin with the potential in the off state, we have the 
 initial conditons as $P_{\text{off}}(x,t=0)=\delta(x)$, and $P_{\text{on}}(x,t=0)=0$. 
 With the help of these initial conditions, Equations (\ref{fpoffFL}) 
 and (\ref{fponFL}) become 
% 
 %%%%%%%%%%%%%%%%%%%%%%%%%%%%%%%%%%%%%%%%%%%%%%%%% 
 \begin{eqnarray} 
 \label{fpoffFLic}
\hspace{-1cm} s\widetilde{P}_{\text{off}}(k,s)-1 = 
-Dk^2\widetilde{P}_{\text{off}}(k,s) 
-\gamma\widetilde{P}_{\text{off}}(k,s)
+\gamma\widetilde{P}_{\text{on}}(k,s), \\ 
 \label{fponFLic}
\hspace{-1cm} s\widetilde{P}_{\text{on}}(k,s) = 
 -Dk^2\widetilde{P}_{\text{on}}(k,s) 
-\mu_0 k\frac{\partial\widetilde{P}_{\text{on}}(k,s)}{\partial k}
-\gamma\widetilde{P}_{\text{on}}(k,s)+\gamma\widetilde{P}_{\text{off}}(k,s). 
 \end{eqnarray} 
 %%%%%%%%%%%%%%%%%%%%%%%%%%%%%%%%%%%%%%%%%%%%%%%%% 
 % 
 Solving for $\widetilde{P}_{\text{off}}$ from Equation (\ref{fpoffFLic}) 
as 
 % 
  %%%%%%%%%%%%%%%%%%%%%%%%%%%%%%%%%%%%%%%%%%%%%%%%% 
 \begin{eqnarray}  
 \label{fpoffFLsol1}
\widetilde{P}_{\text{off}}(k,s) &=& 
\frac{\gamma}{Dk^2+s+\gamma}\widetilde{P}_{\text{on}}(k,s) 
+\frac{1}{Dk^2+s+\gamma}, 
\end{eqnarray} 
 %%%%%%%%%%%%%%%%%%%%%%%%%%%%%%%%%%%%%%%%%%%%%%%%% 
 % 
  and substituting its value in Equation (\ref{fponFLic}) we obtain 
 % 
  %%%%%%%%%%%%%%%%%%%%%%%%%%%%%%%%%%%%%%%%%%%%%%%%% 
 \begin{eqnarray} 
 \label{fponFLsol1}
\hspace{-2.5cm} \frac{\partial\widetilde{P}_{\text{on}}(k,s)}{\partial k} &=& 
 -\frac{1}{\mu_0 k}\left(Dk^2+s+\gamma-\frac{\gamma^2}{Dk^2+s+\gamma}\right)
 \widetilde{P}_{\text{on}}(k,s) 
+\frac{\gamma}{\mu_0 k(Dk^2+s+\gamma)}. 
 \end{eqnarray} 
 %%%%%%%%%%%%%%%%%%%%%%%%%%%%%%%%%%%%%%%%%%%%%%%%% 
 % 
 Defining 
%%%%%%%%%%%%%%%%%%%%%%%%%%%%%%%%%%%%%%%%%%%%%%%%% 
 \begin{eqnarray} 
 \label{funffung} 
\hspace{-2.7cm} f(k,s) = -\frac{1}{\mu_0 k}
 \left(Dk^2+s+\gamma-\frac{\gamma^2}{Dk^2+s+\gamma}\right),
 ~\mathrm{and}~
 g(k,s)=\frac{\gamma}{\mu_0 k(Dk^2+s+\gamma)}, 
 \end{eqnarray} 
%%%%%%%%%%%%%%%%%%%%%%%%%%%%%%%%%%%%%%%%%%%%%%%%%  
we see that $\widetilde{P}_{\text{on}}$ can be solved from Equation 
(\ref{fponFLsol1}), in the Fourier-Laplace domain, as 
 %%%%%%%%%%%%%%%%%%%%%%%%%%%%%%%%%%%%%%%%%%%%%%%%% 
 \begin{eqnarray} 
 \label{fponFLsol2}
\hspace{-1cm} \widetilde{P}_{\text{on}}(k,s) = 
 \widetilde{P}_{\text{on}}(0,s)e^{\int_0^kf(k_1,s)dk_1} 
 +e^{\int_0^kf(k_1,s)dk_1}\int_0^kg(k_1,s)e^{-\int_0^{k_1}f(k_2,s)dk_2}, 
 \end{eqnarray} 
 %%%%%%%%%%%%%%%%%%%%%%%%%%%%%%%%%%%%%%%%%%%%%%%%% 
 % 
 where $k_1$ and $k_2$ are introduced just as the dummy variables of 
 integrations with respect to the Fourier variable. Since the total 
 probability density is $P(x,t)=P_{\text{off}}(x,t)+P_{\text{on}}(x,t)$, which again 
 implies $\widetilde{P}(k,s)=\widetilde{P}_{\text{off}}(k,s)+\widetilde{P}_{\text{on}}(k,s)$ 
 in the Fourier-Laplace space, we obtain with the help of Equations 
 (\ref{fpoffFLsol1}) and (\ref{fponFLsol2}) that 
  %%%%%%%%%%%%%%%%%%%%%%%%%%%%%%%%%%%%%%%%%%%%%%%%% 
 \begin{eqnarray} 
 \label{ptotFLsol}
  \nonumber 
\hspace{-2cm} \widetilde{P}(k,s) = \left( \widetilde{P}_{\text{on}}(0,s)e^{\int_0^kf(k_1,s)dk_1} 
+e^{\int_0^kf(k_1,s)dk_1}\int_0^kg(k_1,s)e^{-\int_0^{k_1}f(k_2,s)dk_2}dk_1 \right) \\ 
\left(1+\frac{\gamma}{Dk^2+s+\gamma}\right)+\frac{1}{Dk^2+s+\gamma}. 
 \end{eqnarray} 
 %%%%%%%%%%%%%%%%%%%%%%%%%%%%%%%%%%%%%%%%%%%%%%%%%
 %
 Inverting the expression of $\widetilde{P}(k,s)$ in 
 Equation (\ref{ptotFLsol}) to get its value in the $(x,t)$ 
 domain is a highly nontrivial task. Therefore, we choose an 
 alternate strategy to calculate its second moment without 
 inverting $\widetilde{P}(k,s)$.  
 From the theory of characteristic function we know that if in the 
 limit $k\to0$ the second derivative of the characteristic function 
 has a finite value, then the modulus of this limiting value is 
 equal to the second moment of the distribution. With this fact 
 in mind we first intend to calculate the second derivative of 
 $\widetilde{P}(k,s)$, which in the limit $k\to0$ will give us 
 the second moment in the Laplace domain. 
 
 Using Equation (\ref{fpoffFLsol1}) we obtain 
%
%%%%%%%%%%%%%%%%%%%%%%%%%%%%%%%%%%%%%%%%%%%%%%%%% 
 \begin{eqnarray} 
 \label{ptotFL} 
 \nonumber 
\widetilde{P}(k,s)&=&\widetilde{P}_{\text{off}}(k,s)+\widetilde{P}_{\text{on}}(k,s) \\ 
&=& \widetilde{P}_{\text{on}}(k,s)\left(1+\frac{\gamma}{Dk^2+s+\gamma}\right)
+\frac{1}{Dk^2+s+\gamma}. 
 \end{eqnarray} 
%%%%%%%%%%%%%%%%%%%%%%%%%%%%%%%%%%%%%%%%%%%%%%%%%
%
Taking limit $k\to0$ on both sides of the Equation (\ref{ptotFL}), 
and using the continuity properties of the characteristic functions, 
we obtain 
%
%%%%%%%%%%%%%%%%%%%%%%%%%%%%%%%%%%%%%%%%%%%%%%%%% 
 \begin{eqnarray} 
 \label{ptotFLk0lim} 
\widetilde{P}(0,s)=\widetilde{P}_{\text{on}}(0,s)\left(1+\frac{\gamma}{s+\gamma}\right)
+\frac{1}{s+\gamma}. 
 \end{eqnarray} 
%%%%%%%%%%%%%%%%%%%%%%%%%%%%%%%%%%%%%%%%%%%%%%%%%
% 
Since $P(x,t)$ is the total probability density which integrates to $1$ 
on the whole real line, the limit of its charateristic function 
$\widehat{P}(k,t)$, as $k\to0$, is $1$. Therefore, we can write 
from Equation (\ref{ptotFLk0lim})  
%
%%%%%%%%%%%%%%%%%%%%%%%%%%%%%%%%%%%%%%%%%%%%%%%%% 
 \begin{eqnarray} 
 \label{ptotFLk0lim1} 
\lim_{k\to0}\int_0^\infty\widehat{P}(k,t)~e^{-st}~dt
=\widetilde{P}_{\text{on}}(0,s)\left(1+\frac{\gamma}{s+\gamma}\right)
+\frac{1}{s+\gamma}. 
 \end{eqnarray} 
%%%%%%%%%%%%%%%%%%%%%%%%%%%%%%%%%%%%%%%%%%%%%%%%%
% 
Taking the limit inside the integral on the LHS of Equation 
(\ref{ptotFLk0lim1}), which is permitted by the Dominated 
Convergence Theorem (DCT), we obtain 
%
%%%%%%%%%%%%%%%%%%%%%%%%%%%%%%%%%%%%%%%%%%%%%%%%% 
 \begin{eqnarray} 
 \label{ptotFLk0lim2} 
 \nonumber 
&&\int_0^\infty1~e^{-st}~dt=\widetilde{P}_{\text{on}}(0,s)\left(1+\frac{\gamma}{s+\gamma}\right)
+\frac{1}{s+\gamma}, \\ 
 \nonumber 
&\Rightarrow&\frac{1}{s}=\widetilde{P}_{\text{on}}(0,s)\left(1+\frac{\gamma}{s+\gamma}\right)
+\frac{1}{s+\gamma}, \\ 
&\Rightarrow& \widetilde{P}_{\text{on}}(0,s)=\frac{\gamma}{s(s+2\gamma)}. 
 \end{eqnarray} 
%%%%%%%%%%%%%%%%%%%%%%%%%%%%%%%%%%%%%%%%%%%%%%%%%
% 
From Equation (\ref{ptotFL}) we calculate the second derivative 
of $\widetilde{P}(k,s)$ with respect to $k$ as 
%
%%%%%%%%%%%%%%%%%%%%%%%%%%%%%%%%%%%%%%%%%%%%%%%%% 
 \begin{eqnarray} 
 \label{der2ptotFL} 
 \nonumber 
\hspace{-2.5cm} \frac{\partial^2\widetilde{P}(k,s)}{\partial k^2} = 
 \gamma\widetilde{P}_{\text{on}}(k,s) \left(\frac{8 D^2 k^2}{\left(\gamma + D 
k^2+s\right)^3}-\frac{2 D}{\left(\gamma +D k^2+s\right)^2}\right)
-\frac{4 \gamma D k }{\left(\gamma +D k^2+s\right)^2}
\frac{\partial\widetilde{P}_{\text{on}}(k,s)}{\partial k} \\ 
\nonumber 
+\frac{\partial^2\widetilde{P}_{\text{on}}(k,s)}{\partial k^2} 
\left(\frac{\gamma }{\gamma +D k^2+s}+1\right)
+\frac{8 D^2 k^2}{\left(\gamma +D k^2+s\right)^3}
-\frac{2D}{\left(\gamma +D k^2+s\right)^2}. 
 \end{eqnarray} 
%%%%%%%%%%%%%%%%%%%%%%%%%%%%%%%%%%%%%%%%%%%%%%%%%
% 
Therefore, in the limit $k\to0$ we obtain 
%
%%%%%%%%%%%%%%%%%%%%%%%%%%%%%%%%%%%%%%%%%%%%%%%%% 
 \begin{eqnarray} 
 \label{der2ptotFLk01} 
 \nonumber 
 \lim_{k\to0}\frac{\partial^2\widetilde{P}(k,s)}{\partial k^2}=
-\frac{2\gamma D}{(\gamma +s)^2}
\left(\lim_{k\to0}\widetilde{P}_{\text{on}}(k,s)\right) \\
+\left(\frac{\gamma }{\gamma+s}+1\right) 
\left(\lim_{k\to0}\frac{\partial^2\widetilde{P}_{\text{on}}(0,s)}
{\partial k^2}\right) 
-\frac{2D}{(\gamma+s)^2}, 
 \end{eqnarray} 
%%%%%%%%%%%%%%%%%%%%%%%%%%%%%%%%%%%%%%%%%%%%%%%%%
% 
provided the limits $\mathrm{lim}_{k\to0}\widetilde{P}_{\text{on}}(k,s)$ 
and $\mathrm{lim}_{k\to0}\partial_k^2\widetilde{P}_{\text{on}}(k,s)$ exist. 

We now show that both the above mentioned limits exist. 
Representing the RHS of (\ref{fponFLsol1}) as a fraction 
we observe that 
%
%%%%%%%%%%%%%%%%%%%%%%%%%%%%%%%%%%%%%%%%%%%%%%%%% 
 \begin{eqnarray} 
 \label{der1p1} 
 \frac{\partial \widetilde{P}_{\text{on}}(k,s)}{\partial k}= 
 \frac{\gamma -\left(D k^2+s\right)\left(2 \gamma 
 +Dk^2+s\right)\widetilde{P}_{\text{on}}(k,s)}
{k \mu_0  \left(\gamma +D k^2+s\right)}, 
 \end{eqnarray} 
%%%%%%%%%%%%%%%%%%%%%%%%%%%%%%%%%%%%%%%%%%%%%%%%%
% 
If we take limit $k\to0$ on the both sides of Equation (\ref{der1p1}), 
and use the value of $\widetilde{P}_{\text{on}}(0,s)$ from Equation 
(\ref{ptotFLk0lim2}), we observe that the numerator and the 
denominator of the RHS both tend to zero, and hence, it is required 
to apply the L'Hospital's rule on the RHS to evaluate this limit. 
After applying L'Hospital's rule on the RHS of Equation (\ref{der1p1}) 
we obtain 
%
%%%%%%%%%%%%%%%%%%%%%%%%%%%%%%%%%%%%%%%%%%%%%%%%% 
 \begin{eqnarray} 
 \label{der1p1k0} 
 \lim_{k\to0}\frac{\partial \widetilde{P}_{\text{on}}(k,s)}{\partial k}= 
 \frac{s (2 \gamma +s)}{\mu_0  (\gamma +s)}
 \left(\lim_{k\to0}
 \frac{\partial \widetilde{P}_{\text{on}}(k,s)}{\partial k}\right). 
 \end{eqnarray} 
%%%%%%%%%%%%%%%%%%%%%%%%%%%%%%%%%%%%%%%%%%%%%%%%%
% 
Solving Equation (\ref{der1p1k0}) for 
$\mathrm{lim}_{k\to0}\partial_k\widetilde{P}_{\text{on}}(k,s)$, 
we obtain $\mathrm{lim}_{k\to0}\partial_k\widetilde{P}_{\text{on}}(k,s)=0$. 
To evaluate $\mathrm{lim}_{k\to0}\partial_k^2\widetilde{P}_{\text{on}}(k,s)$ 
we take derivative on the both sides of the Equation (\ref{der1p1}) 
with respect to $k$, and take the limit $k\to0$ to obtain 
%
%%%%%%%%%%%%%%%%%%%%%%%%%%%%%%%%%%%%%%%%%%%%%%%%% 
 \begin{eqnarray} 
 \label{abcd} 
\lim_{k\to0}\frac{\partial^2\widetilde{P}_{\text{on}}(k,s)}{\partial k^2}= 
\lim_{k\to0}\frac{Q(k,s)}{k^2 \mu_0\left(\gamma +Dk^2+s\right)^2},  
 \end{eqnarray} 
%%%%%%%%%%%%%%%%%%%%%%%%%%%%%%%%%%%%%%%%%%%%%%%%%
% 
where
%
%%%%%%%%%%%%%%%%%%%%%%%%%%%%%%%%%%%%%%%%%%%%%%%%% 
 \begin{eqnarray} 
 \label{der1p1numer} 
\nonumber 
\hspace{-2.5cm} Q(k,s)=\widetilde{P}_{\text{on}}(k,s) \left(-D^3 k^6-D^2 k^4 (\gamma +s) 
+D k^2 \left(-2 \gamma^2+s^2+2 \gamma  s\right)+s (\gamma +s) (2 \gamma +s)\right) \\ 
\nonumber 
\hspace{-2cm} -k \left(Dk^2+s\right) \frac{\partial\widetilde{P}_{\text{on}}(k,s)}{\partial k} \left(\gamma +D k^2+s\right) \left(2
   \gamma +D k^2+s\right)-\gamma  \left(\gamma +3 D k^2+s\right). 
 \end{eqnarray} 
%%%%%%%%%%%%%%%%%%%%%%%%%%%%%%%%%%%%%%%%%%%%%%%%%
% 
Again, we observe that in the limit $k\to0$, both the numerator and 
the denominator on the RHS of the Equation (\ref{abcd}) tend to 
zero (using the vales of $\mathrm{lim}_{k\to0}\widetilde{P}_{\text{on}}(k,s)$ 
and $\mathrm{lim}_{k\to0}\partial_k\widetilde{P}_{\text{on}}(k,s)$ from 
Equations (\ref{ptotFLk0lim2}) and (\ref{der1p1k0}), respectively). 
This demands another application of L'Hospital's rule in evaluating 
the limit in Equation (\ref{der1p1k0}). Therefore, applying 
L'Hospital's rule to the RHS of Equation (\ref{abcd}), 
and using Equations (\ref{ptotFLk0lim2}) and (\ref{der1p1k0}), 
we obtain 
%
%%%%%%%%%%%%%%%%%%%%%%%%%%%%%%%%%%%%%%%%%%%%%%%%% 
 \begin{eqnarray} 
 \label{der2p1k02} 
\hspace{-2.5cm}\lim_{k\to0}\frac{\partial^2\widetilde{P}_{\text{on}}(k,s)}{\partial k^2}= 
-\frac{6\gamma D+\frac{2\gamma D\left(2\gamma^2-s^2-2\gamma 
s\right)}{s (2\gamma +s)}
+\left(\lim_{k\to0}\frac{\partial^2\widetilde{P}_{\text{on}}(k,s)}{\partial k^2}\right)s(\gamma+s)(2\gamma +s)}{2 \mu_0  (\gamma +s)^2}.   
 \end{eqnarray} 
%%%%%%%%%%%%%%%%%%%%%%%%%%%%%%%%%%%%%%%%%%%%%%%%%
% 
Solving for $\mathrm{lim}_{k\to0}\partial_k^2\widetilde{P}_{\text{on}}(k,s)$ 
from Equation (\ref{der2p1k02}), and substituting its value in 
Equation (\ref{der2ptotFLk01}), we get 
%
%%%%%%%%%%%%%%%%%%%%%%%%%%%%%%%%%%%%%%%%%%%%%%%%% 
 \begin{eqnarray} 
 \label{varLap}  
 \lim_{k\to0}\frac{\partial^2\widetilde{P}(k,s)}{\partial k^2}=
-\frac{2D}{s}
\left(\frac{2\gamma}{2\gamma\mu_0+s^2+2s(\gamma+\mu_0)}
+\frac{1}{2\gamma+s}\right), 
 \end{eqnarray} 
%%%%%%%%%%%%%%%%%%%%%%%%%%%%%%%%%%%%%%%%%%%%%%%%%
% 
which is the Laplace transform of the negative of the second moment 
of the position distribution. Therefore, inverting this expression 
and taking its absolute value, we obtain the time dependent variance 
of the distribution as 
%
%%%%%%%%%%%%%%%%%%%%%%%%%%%%%%%%%%%%%%%%%%%%%%%%% 
 \begin{eqnarray} 
 \label{vartime} 
 \nonumber 
\hspace{-2cm} \sigma_{\text{off}}^2(t)=
\frac{D(2\gamma+\mu_0)}{\gamma\mu_0}-\frac{De^{-2\gamma t}}{\gamma} 
\nonumber 
-~\frac{D}{\mu_0\sqrt{\gamma^2+\mu_0^2}}
\left\{-\gamma e^{t
   \left(-\sqrt{\gamma ^2+\mu_0 ^2}-\gamma -\mu_0 \right)}  \right. \\ \left.
   \nonumber
   +\gamma  e^{t
   \left(\sqrt{\gamma ^2+\mu_0 ^2}-\gamma -\mu_0 \right)}  -\mu_0  e^{t \left(-\sqrt{\gamma^2+\mu_0 ^2}-\gamma -\mu_0 \right)} 
   +\mu_0 e^{t\left(\sqrt{\gamma^2+\mu_0^2}-\gamma-\mu_0 \right)} \right\} \\   
-\frac{D}{\mu_0}\left\{e^{t\left(-\sqrt{\gamma^2+\mu_0^2}-\gamma-\mu_0\right)}
   +e^{t\left(\sqrt{\gamma^2+\mu_0^2}-\gamma-\mu_0\right)}\right\},  
 \end{eqnarray} 
%%%%%%%%%%%%%%%%%%%%%%%%%%%%%%%%%%%%%%%%%%%%%%%%%
% 
where the suffix ``off'' is to emphasize on the fact that the 
potential is in the ``off'' state at $t=0$. 
Since $(\gamma+\mu_0) \geq \sqrt{\gamma^2+\mu_0^2}$, we observe from 
Equation (\ref{vartime}) that in the steady sate, i.e., when 
$t\to\infty$, the variance becomes 
%
%
%%%%%%%%%%%%%%%%%%%%%%%%%%%%%%%%%%%%%%%%%%%%%%%%% 
 \begin{eqnarray} 
 \label{varsteady} 
 \sigma_{\text{off}}^2(\infty)=D\left(\frac{2}{\mu_0}+\frac{1}{\gamma}\right). 
 \end{eqnarray} 
%%%%%%%%%%%%%%%%%%%%%%%%%%%%%%%%%%%%%%%%%%%%%%%%%

Also, we see in Equation (\ref{vartime}) that there are three 
time scales in the expression of the time-dependent variance, 
which are $\tau_1=1/\gamma$, $\tau_2=1/(\gamma+\mu_0+\sqrt{\gamma^2+\mu_0^2})$, 
and $\tau_3=1/(\gamma+\mu_0-\sqrt{\gamma^2+\mu_0^2})$. It is evident that 
$\tau_3 > \tau_2$. Since $0 \geq (\mu_0-\sqrt{\gamma^2+\mu_0^2})$, 
and hence, $\gamma \geq (\gamma+\mu_0-\sqrt{\gamma^2+\mu_0^2})$, we 
conclude that $\tau_3 \geq \tau_1$. Therefore, $\tau_3=1/(\gamma+\mu_0-\sqrt{\gamma^2+\mu_0^2})$ is the largest time scale in this dynamical 
system. 

The time-dependent variance $\sigma_{\text{off}}(t)$ in Equation (\ref{vartime}) 
is obtained in the case when at time $t=0$ the potential is in the 
``off'' state. On the other hand, if we start with the potential 
in the ``on'' state, we also get the variance $\sigma_{\text{on}}(t)$, 
just by following the similar procedure. In that case, 
using the initial conditions $\widehat{P}_{\text{off}}(k,0)=0$ and 
$\widehat{P}_{\text{on}}(k,0)=1$ in Equations (\ref{fpoffFL}) and 
(\ref{fponFL}), respectively, and subsequently, following the 
same procedure as above we get the time dependent variance 
$\sigma^2_{\text{on}}(t)$ as 
%
%%%%%%%%%%%%%%%%%%%%%%%%%%%%%%%%%%%%%%%%%%%%%%%%% 
 \begin{eqnarray} 
 \label{vartimerev} 
 \nonumber 
 \sigma_{\text{on}}^2(t)&=& 
 \frac{D(2\gamma+\mu_0)}{\gamma\mu_0}+\frac{D e^{-2\gamma t}}{\gamma } \\ 
 \nonumber 
 && -\frac{D}{\gamma \mu_0 \sqrt{\gamma ^2+\mu_0 ^2}} 
 \left\{-\gamma^2 e^{t\left(-\sqrt{\gamma^2+\mu_0^2}-\gamma-\mu_0
   \right)}+\gamma^2 e^{t \left(\sqrt{\gamma ^2+\mu_0 ^2}-\gamma -\mu_0
   \right)} \right. \\ 
\nonumber 
&& \left. -\mu_0^2 e^{t \left(-\sqrt{\gamma ^2+\mu_0 ^2}-\gamma -\mu_0 \right)}+\mu_0 ^2
   e^{t \left(\sqrt{\gamma ^2+\mu_0 ^2}-\gamma -\mu_0 \right)}\right\} \\ 
\nonumber 
&& -\frac{D}{\gamma\mu_0}
 \left\{\gamma e^{t\left(-\sqrt{\gamma^2+\mu_0^2}-\gamma-\mu_0\right)}
 +\gamma e^{t \left(\sqrt{\gamma^2+\mu_0^2}-\gamma-\mu_0\right)} \right. \\ 
%\nonumber 
&& \left. +\mu_0 e^{t\left(-\sqrt{\gamma^2+\mu_0^2}-\gamma-\mu_0\right)}+\mu_0 e^{t
 \left(\sqrt{\gamma^2+\mu_0^2}-\gamma-\mu_0\right)}\right\}.   
 \end{eqnarray} 
%%%%%%%%%%%%%%%%%%%%%%%%%%%%%%%%%%%%%%%%%%%%%%%%%
%
\section{Discussion for $\gamma_{\text{on}} \ne \gamma_{\text{off}}$}
\label{different_rates}

In this section we discuss the stationary distribution for different switching on and off rates ($\gamma_{\text{on}}$ and $\gamma_{\text{off}}$, respectively)---when in on-state the potential is turned off at a hazard rate $\gamma_{\text{on}}$, while in off-state the potential is turned on at the rate $\gamma_{\text{off}}$. Thus the Fokker-Planck equations \eref{eq:fp_ss_on} and \eref{eq:fp_ss_of} have the form
\bea
\mu_0\frac{\partial }{\partial x}(x\pon(x))+D\frac{\partial^2 \pon (x)}{\partial x^2}-\gamma_{\text{on}} \pon(x)+\gamma_{\text{off}} \pof(x)&=&0,\\
D\frac{\partial^2 \pof{(x)}}{\partial x^2}-\gamma_{\text{off}} \pof(x)+\gamma_{\text{on}} \pon(x)&=&0.
\eea
Following a similar procedure as described in \sref{stationary_distribution}, here we find the solution in Fourier space as
\bea
\hat{P}_{\text{on}}(k)&=&  \left(\frac{\gamma_{\text{off}}}{\gamma_{\text{on}} + \gamma_{\text{off}}} \right) \frac{e^{-Dk^2/(2\mu_0)}}{[f_1(k)]^{\gamma_{\text{on}}/(2\mu_0)}}, 
\eea
\bea
\hat{P}_{\text{off}}(k)&=& \left(\frac{\gamma_{\text{on}}}{\gamma_{\text{on}} + \gamma_{\text{off}}} \right) \frac{e^{-Dk^2/(2\mu_0)}}{[f_1(k)]^{1 + \gamma_{\text{on}}/(2\mu_0)}},
\eea
where $f_1(k)=(1+Dk^2/\gamma_{\text{off}})$ and $\hat{P}(k) = \hat{P}_{\text{on}}(k) + \hat{P}_{\text{off}}(k)$. 
The generalization of Eqs. \eref{g1}-\eref{Px_genn} follows trivially. 
In particular, the corresponding form of Eq.~\eref{eq:refract} is 
\bea
P(x)&=&\left(\frac {\gamma_{\text{off}}}{\gamma_{\text{on}}+\gamma_{\text{off}}}\right)\delta(x)+\left(\frac {\gamma_{\text{on}}}{\gamma_{\text{on}}+\gamma_{\text{off}}}\right)\frac{\sqrt{\gamma_{\text{off}}/D}}{2}e^{-\sqrt{\gamma_{\text{off}}/D}\,|x|}.
\eea
In the limit $\gamma_{\text{off}}/\gamma_{\text{on}}\ll 1$, the first term on the rhs of the above vanishes and we obtain,
\bea
P(x)\approx\frac{\sqrt{\gamma_{\text{off}}/D}}{2}e^{-\sqrt{\gamma_{\text{off}}/D}|x|}.
\eea
This is exactly the result of a diffusion with instantaneous resetting without refractory period \cite{evans2011diffusion}. Physically, as soon as the trap is turned on, the particle still returns  to the origin instantaneously (as the trap strength is very large), however as $\gamma_{\text{off}}/\gamma_{\text{on}}\ll 1$ the trap is turned off very fast. Thus, there is no refractory period after a resetting event.

\section{Mean First Passage Time for an Ornstein-Uhlenbeck process} 
\label{MFPT_ou}
We provide the derivation of the mean first-passage time to an absorbing target for the Ornstein-Uhlenbeck process used in Eq.~\eref{MFPT_ou_m} in the main text for the sake of completeness. Let $Q(x_0,t)$ denote the survival probability i.e., the probability that a Brownian particle in a harmonic trap ($V(x)=\mu_0 x^2/4$) does not reach a target located at $x=x_{\text{abs}}$ at time $t$, starting from $x(t=0)=x_0$. We can immediately write the backward Fokker Planck equation for the survival probability,
\bea
\frac{\partial Q(x_0,t)}{\partial t}=-\frac{\mu_0}{2}x_0\frac{\partial Q(x_0,t)}{\partial x_0}+D\frac{\partial^2 Q(x_0,t)}{\partial x^2}.
\label{backfp}
\eea
We need to solve this equation with the initial condition, $Q(x_0,0)=1$ and the boundary conditions $Q(\infty,t)=1$ and $Q(x_{\text{abs}},t)=0$.
We take a Laplace transform of the Eq.~\eref{backfp} with respect to $t$ defined by $\tilde{f}(s)=\int_{0}^{\infty}e^{-st}f(t) dt$ followed by a variable transformation, $G(x_0,s)=\tilde{Q}(x_0,s)-1/s$ to get,
\bea
D \frac{\partial^2 G(x_0,s)}{\partial x^2}-\frac{\mu_0}{2}x\frac{\partial G(x_0,s)}{\partial x}-s G(x_0,s)=0
\eea
 Now, substituting 
 \bea
 G(x_0,s)=e^{\mu_0 x^2/(8D)}W(x_0\sqrt{\mu_0/2D}),
\eea
we get,
\bea
W''(z)+(-2s/\mu_0+1/2-z^2/4)W(z)=0
\eea
where $z=x_0\sqrt{\mu_0/2D}$. The general solution of the above equation can be written in terms of the Parabolic cylinder functions $\cal{D}_{-2s/\mu_0}(x_0\sqrt{\mu_0/2D})$ and $\cal{D}_{-2s/\mu_0}(-x_0\sqrt{\mu_0/2D})$. Noting that $\cal{D}_{-2s/\mu_0}(x_0\sqrt{\mu_0/2D})\to\infty$ and $\cal{D}_{-2s/\mu_0}(-x_0\sqrt{\mu_0/2D})\to 0$ as $x_0\to\infty$, we conclude
\bea
G(x_0,s)=A e^{\mu_0 x^2/(8D)}\cal{D}_{-2s/\mu_0}(-x_0\sqrt{\mu_0/2D}).
\eea
Using the absorbing boundary condition at $x =x_{abs}$, we obtain
\bea
A=e^{-\mu_0 x_{\text{abs}}^2/(8D)}\cal{D}_{-2s/\mu_0}(-x_{\text{abs}}\sqrt{\mu_0/2D}).
\eea
Thus the survival probability in the Laplace space comes out to be
\bea
\tilde{Q}(x_0,s)=\frac{1}{s}\left(1-e^{-\mu_0(x_0^2- x_{\text{abs}}^2)/(8D)}\frac{\cal{D}_{-2s/\mu_0}(-x_0\sqrt{\mu_0/2D})}{\cal{D}_{-2s/\mu_0}(-x_{\text{abs}}\sqrt{\mu_0/2D})}\right).
\label{ou:mfpt}
\eea
%The survival probability is the cumulative first-passage probability $F(t)$, i.e., $-\frac{\partial Q(x_0,t)}{\partial t}$. 
%Using this the mean first passage time (MFPT) is,
%\bea
%\text{MFPT}=\int_{0}^{\infty}t' F(x_0,t') dt'=\int_{0}^{\infty}Q(x_0,t')dt'
%\eea
The mean first passage time (MFPT) can be obatined by taking $s \rightarrow 0$ of Eq. \eref{ou:mfpt}~\cite{redner2001}, which we have used in Eq. \eref{MFPT_ou_m} in the main text.
%Thus, MFPT$=\tilde{Q}(x_0,s\to 0)$. This combined with Eq.~\eref{ou:mfpt} is used in the main text.
%--------------------------------------------------------------------
\section{Mean first passage time in resetting with refractory period}
\label{mfpt_ref}

In this section, we discuss the behavior of the mean first passage time, in the limit where the strength of the harmonic potential $(\mu_0)$ is large.
In this limit, the particle quickly returns to the centre of the harmonic potential as soon as the potential is turned on.
In fact, the dynamics of the particle in the limit $\mu_0 \rightarrow \infty$ becomes equivalent to the dynamics of a Brownian particle undergoing a stochastic resetting followed by a random refractory period during which the particle does not move \cite{refractoryperiod}. 
In the presence of refractory periods the mean first passage time at $x_{abs} < 0$, starting from the origin $(x = 0)$, can be written as \cite{refractoryperiod}
\bea 
T(x_{abs}) = \frac{\int_0^{\infty} dt \;\left[ g(t) + \int_0^{\infty} d\tau \; \tau \; H(t,\tau)\right]\;Q_0(x_{abs},t)}{1 - \int_0^{\infty} dt \;h(t)\;Q_0(x_{abs},t)},
\label{T_x0}
\eea
where $H(t, \tau)$ is the joint probability density of resetting to be occurred at time $t$ followed by a refractory interval of time $\tau$, $h(t)$ is the marginal distribution of time $t$ of the reset event given by $h(t) = \int_0^{\infty} d\tau\; H(t, \tau)$, $g(t) = \int_{t}^{\infty} dt' \;h(t')$ denotes the corresponding survival probability, and $Q_0(x_{abs},t)$ represents the survival probability at $x_{abs} < 0$ starting from the origin in the absence of resetting.
In the absence of resetting, the particle in our case is execuites an overdamped Brownian motion. 
In this case the survival probability $Q_0(x_{abs},t) = \text{Erf}[|x_{abs}|/\sqrt{4 D t}]$.
Note that in the present case both the resetting and refractroy period occur at the rate $\gamma$, and hence, $H(t, \tau) = \gamma^2 e^{-\gamma(t + \tau)}$, $h(t) = \gamma e^{-\gamma t}$, and $g(t) = e^{-\gamma t}$.
Using these results in \eref{T_x0} we obtain
\bea 
T(x_{abs}) = \frac{2}{\gamma} \left( e^{\sqrt{\gamma/D}\;|x_{abs}|} - 1\right),
\eea 
which we directly use in the main text as the expression of MFPT$_{\text{ref}}$ in Eq. \eref{MFPT_rte}.\\
%--------------------------------------------------------------------
 

\vspace*{0.5 cm}
\section*{References}
\begin{thebibliography}{99}

\bibitem{brown1} S. Chandrasekhar, Rev. Mod. Phys. 15, 1 (1943).

\bibitem{brown2} B. Duplantier, Progress in Mathematical Physics, vol 47. Birkhäuser Basel (2006).

\bibitem{brown3} E. Frey and K. Kroy, Ann. Phys. (Leipzig)
14, 20 (2005)

\bibitem{eco} J. G. Skellam, Biometrika 38, 196 (1951).

\bibitem{comp} S. N. Majumdar, Current Science 89, 2076 (2005).

\bibitem{finance} P. H. Cootner, Ed., (MIT press, Cambridge, Massachusetts,1964).

\bibitem{ou} Risken, H. (1984), Springer-Verlag, pp. 99–100, ISBN 978-0-387-13098-9

\bibitem{evans2011diffusion}
M.~R. Evans and S.~N. Majumdar, {
  Physical review letters}, vol.~106, no.~16, p.~160601, 2011.

\bibitem{evans2020stochastic}
M.~R. Evans, S.~N. Majumdar, and G.~Schehr, { Journal of Physics A: Mathematical and Theoretical},
  vol.~53, no.~19, p.~193001, 2020.

\bibitem{evans2011diffusion2}
M.~R. Evans and S.~N. Majumdar,'' {
  Journal of Physics A: Mathematical and Theoretical}, vol.~44, no.~43,
  p.~435001, 2011.

%\bibitem{transition}S. N. Majumdar, S. Sabhapandit and G. Schehr,``Dynamical transition in the temporal relaxation of stochastic processes under resetting'' Phys. Rev. E 91, 052131 (2015).


%\bibitem{nagar2016diffusion}
%A.~Nagar and S.~Gupta, ``Diffusion with stochastic resetting at power-law
%  times,'' {\em Physical Review E}, vol.~93, no.~6, p.~060102, 2016.

%\bibitem{eule2016non}
%S.~Eule and J.~J. Metzger, ``Non-equilibrium steady states of stochastic
%  processes with intermittent resetting,'' {  New Journal of Physics},
%  vol.~18, no.~3, p.~033006, 2016.


\bibitem{montero2013monotonic}
M.~Montero and J.~Villarroel, {  Physical Review E}, vol.~87, no.~1, p.~012116, 2013.

\bibitem{pal2015diffusion} A. Pal, Phys. Rev. E 91, 012113 (2015).
%Diffusion in a potential landscape with stochastic resetting (2015)

\bibitem{gupta2019stochastic} D. Gupta, J. Stat. Mech. 033212 (2019). 

\bibitem{shkilev2017continuous}
V.~Shkilev,  { Physical Review E}, vol.~96, no.~1, p.~012126, 2017.

\bibitem{montero2017continuous}
M.~Montero, A.~Mas{\'o}-Puigdellosas, and J.~Villarroel, {  The European Physical Journal B},
  vol.~90, no.~9, p.~176, 2017.

\bibitem{zhou2020continuous}
T.~Zhou, P.~Xu, and W.~Deng, {  Physical Review Research}, vol.~2, no.~1,
  p.~013103, 2020.

\bibitem{kusmierz2014first}
L.~Kusmierz, S.~N. Majumdar, S.~Sabhapandit, and G.~Schehr, 
  {  Physical review letters}, vol.~113, no.~22, p.~220602, 2014.

\bibitem{kusmierz2015optimal}
{\L}.~Ku{\'s}mierz and E.~Gudowska-Nowak, {  Physical Review E}, vol.~92, no.~5,
  p.~052127, 2015.

\bibitem{convexhull} S. N. Majumdar, F. Mori, H. Schawe, and G. Schehr, Phys. Rev. E 103, 022135.

\bibitem{rap} P. Singh, Journal of Physics A: Mathematical and Theoretical, Volume 53, Number 40, 2020.


\bibitem{resetrtp1}
M.~R. Evans and S.~N. Majumdar, {  Journal of Physics A: Mathematical and Theoretical},
  vol.~51, no.~47, p.~475003, 2018.
  
\bibitem{resetrtp2} I. Santra, U. Basu, S. Sabhapandit, J. Stat. Mech. (2020) 113206

\bibitem{resetabp1} A. Scacchi and A. Sharma, Molecular Physics, 116, 460 (2017).

\bibitem{resetabp2} V. Kumar, O. Sadekar, U. Basu Phys. Rev. E 102, 052129. 


\bibitem{enzyme1} S. Reuveni, Phys. Rev. Lett. 116, 170601 (2016).


\bibitem{enzyme2} S. Reuveni, M. Urbakh, and J. Klafter, Proc. Nat. Acad. Sci. 111, 4391 (2014).


\bibitem{activetransport} P. C. Bressloff, J. Phys. A: Math. Theor. 53, 355001 (2020)


\bibitem{interface1} S. Gupta, S. N. Majumdar, and G. Schehr, Phys. Rev. Lett. 112, 220601 (2014)

\bibitem{interface2} S. Gupta and A. Nagar, J. Phys. A: Math. Theor. 49, 445001 (2016)
\bibitem{reactdiff} X. Durang, M. Henkel, and H. Park, J. Phys. A: Math. Theor. 47, 045002 (2014)


\bibitem{isingreset} M. Magoni, S. N. Majumdar, and G. Schehr, Phys. Rev. Research 2, 033182 (2020)

\bibitem{sepreset1} U. Basu, A. Kundu, and A. Pal, Phys. Rev. E 100, 032136 (2019)
 
\bibitem{sepreset2} S. Karthika and A. Nagar, J. Phys. A: Math. Theor. 53, 115003 (2020)

\bibitem{refractoryperiod} M. R. Evans and S. N. Majumdar, J. Phys. A: Math. Theor. 52 01LT01 (2019).


%\bibitem{masoliver2019telegraphic} J.~Masoliver, ``Telegraphic processes with stochastic resetting,'' {  Physical Review E}, vol.~99, no.~1, p.~012121, 2019.

%\bibitem{bhat2016stochastic}
%U.~Bhat, C.~De~Bacco, and S.~Redner, ``Stochastic search with poisson and
%  deterministic resetting,'' {  Journal of Statistical Mechanics: Theory and Experiment}, vol.~2016, no.~8, p.~083401, 2016.

%\bibitem{pal2017first}
%A.~Pal and S.~Reuveni, ``First passage under restart,'' {  Physical review letters}, vol.~118, no.~3, p.~030603, 2017.

%\bibitem{pal2015diffusion} A.~Pal, ``Diffusion in a potential landscape with stochastic resetting,'' {   Physical Review E}, vol.~91, no.~1, p.~012113, 2015.

%\bibitem{ray2019peclet}
%S.~Ray, D.~Mondal, and S.~Reuveni, ``P{\'e}clet number governs transition to acceleratory restart in drift-diffusion,'' {  Journal of Physics A:
%Mathematical and Theoretical}, vol.~52, no.~25, p.~255002, 2019.

%\bibitem{ahmad2019first} S.~Ahmad, I.~Nayak, A.~Bansal, A.~Nandi, and D.~Das, ``First passage of a particle in a potential under stochastic resetting: A vanishing transition of optimal resetting rate,'' {  Physical Review E}, vol.~99, no.~2, p.~022130, 2019.


%\bibitem{christou2015diffusion}
%C.~Christou and A.~Schadschneider, ``Diffusion with resetting in bounded domains,'' {  Journal of Physics A: Mathematical and Theoretical}, vol.~48, no.~28, p.~285003, 2015.

%\bibitem{chatterjee2018diffusion}
%A.~Chatterjee, C.~Christou, and A.~Schadschneider, ``Diffusion with resetting inside a circle,'' {  Physical Review E}, vol.~97, no.~6, p.~062106, 2018.

%\bibitem{pal2019local} A.~Pal, R.~Chatterjee, S.~Reuveni, and A.~Kundu, ``Local time of diffusion with stochastic resetting,'' {  Journal of Physics A: Mathematical and Theoretical}, vol.~52, no.~26, p.~264002, 2019.

%\bibitem{pal2019first}A.~Pal and V.~Prasad, ``First passage under stochastic resetting in an interval,'' {  Physical Review E}, vol.~99, no.~3, p.~032123, 2019.

%\bibitem{evans2014diffusion} M.~R. Evans and S.~N. Majumdar, ``Diffusion with resetting in arbitrary spatial dimension,'' {  Journal of Physics A: Mathematical and Theoretical}, vol.~47, no.~28, p.~285001, 2014.

%\bibitem{falcao2017interacting} R.~Falcao and M.~R. Evans, ``Interacting brownian motion with resetting,'' {  Journal of Statistical Mechanics: Theory and Experiment}, vol.~2017, no.~2, p.~023204, 2017.

%\bibitem{de2020first}
%B.~De~Bruyne, J.~Randon-Furling, and S.~Redner, ``First-passage resetting and optimization,'' {  arXiv preprint arXiv:2005.00957}, 2020.

%\bibitem{pal2019time} A.~Pal, {\L}.~Ku{\'s}mierz, and S.~Reuveni, ``Time-dependent density of  diffusion with stochastic resetting is invariant to return speed,'' {  Physical Review E}, vol.~100, no.~4, p.~040101, 2019.

%\bibitem{bodrova2020resetting}
%A.~S. Bodrova and I.~M. Sokolov, ``Resetting processes with noninstantaneous return,'' {  Physical Review E}, vol.~101, no.~5, p.~052130, 2020.

\bibitem{besga2020optimal}
B.~Besga, A.~Bovon, A.~Petrosyan, S.~N. Majumdar, and S.~Ciliberto,  {  Physical Review Research},
  vol.~2, no.~3, p.~032029, 2020.

\bibitem{tal2020experimental}
O.~Tal-Friedman, A.~Pal, A.~Sekhon, S.~Reuveni, and Y.~Roichman,  {  The journal of
  physical chemistry letters}, vol.~11, no.~17, pp.~7350--7355, 2020.
   

\bibitem{pal19} A. Pal, L. Kuśmierz and S. Reuveni 2019 Phys. Rev. E 100 040101.

\bibitem{bodrova2020_1}
A.~S. Bodrova and I.~M. Sokolov, {  Physical Review E}, vol.~102, p.~032129, 2020.

\bibitem{bodrova2020_2}
A.~S. Bodrova and I.~M. Sokolov, {  Physical Review E}, vol.~101, p.~052130, 2020.

%Bodrova A.S., Sokolov I.M. Phys. Rev. E 102, 032129 (2020).
%Bodrova A.S., Sokolov I.M. Phys. Rev. E  101, 052130 (2020).

%\bibitem{bodrova2019nonrenewal}
%A.~S. Bodrova, A.~V. Chechkin, and I.~M. Sokolov, {  Physical Review E}, vol.~100, no.~1, p.~012119, 2019.

%\bibitem{bodrova2019scaled} A.~S. Bodrova, A.~V. Chechkin, and I.~M. Sokolov, {  Physical Review E}, vol.~100, no.~1, p.~012120, 2019.

\bibitem{pal20} A. Pal, L. Kuśmierz and S. Reuveni Phys. Rev. Research 2, 043174 (2020).

\bibitem{gupta2020resetting}
D.~Gupta, A.~Pal, and A.~Kundu, {  arXiv preprint arXiv:2012.12878}, 2020.

\bibitem{gupta2020stochastic}
D.~Gupta, C.~A. Plata, A.~Kundu, and A.~Pal,  {  Journal of Physics A:
  Mathematical and Theoretical}, vol.~54, no.~2, p.~025003, 2020.

\bibitem{mercado2020intermittent}
G.~Mercado-V{\'a}squez, D.~Boyer, S.~N. Majumdar, and G.~Schehr,  {  Journal of Statistical Mechanics: Theory and
  Experiment}, vol.~2020, no.~11, p.~113203, 2020.



\bibitem{rtp1} K. Malakar, V. Jemseena, A. Kundu, K. Vijay Kumar, S. Sabha-
pandit, S. N. Majumdar, S. Redner, A. Dhar, JSTAT 043215 (2018).


\bibitem{rtp2} I. Santra, U. Basu, S. Sabhapandit, Phys. Rev. E
101, 062120 (2020).

\bibitem{abp1} A. Pototsky and H. Stark, EPL 98, 50004 (2012).

\bibitem{abp2} U. Basu, S. N. Majumdar, A. Rosso, G. Schehr, Phys. Rev. E 98, 062121 (2018).

\bibitem{drabp} I. Santra, U. Basu, S. Sabhapandit, arXiv preprint,	arXiv:2101.11327 (2021).

\bibitem{ashkin1997optical}
A. Ashkin, {  Proceedings of the National Academy of Sciences}, vol.~94,
  no.~10, pp.~4853--4860, 1997.


\bibitem{datar2015dynamics}
A.~Datar, T.~Bornschl{\"o}gl, P.~Bassereau, J.~Prost, and P.~A. Pullarkat, {  Biophysical journal}, vol.~108, no.~3,
  pp.~489--497, 2015.


\bibitem{redner2001} S. Redner, A guide to first-passage processes, Cambridge university press (2001). 


%
%\bibitem{weron2004modeling}
%R.~Weron, M.~Bierbrauer, and S.~Tr{\"u}ck, ``Modeling electricity prices: jump
%  diffusion and regime switching,'' {  Physica A: Statistical Mechanics and
%  its Applications}, vol.~336, no.~1-2, pp.~39--48, 2004.
%
%\bibitem{chakraborty2012persistence}
%D.~Chakraborty, ``Persistence of a brownian particle in a time-dependent
%  potential,'' {  Physical Review E}, vol.~85, no.~5, p.~051101, 2012.
%
%

\end{thebibliography}
\end{document}